\documentclass[prl,twocolumn,aps]{revtex4}
\usepackage{epsfig}
\newcommand{\be}{\begin{equation}}
\newcommand{\ee}{\end{equation}}
\newcommand{\bea}{\begin{eqnarray}}
\newcommand{\eea}{\end{eqnarray}}

\newcommand{\bwt}{\begin{widetext}}
\newcommand{\ewt}{\end{widetext}}
\newcommand{\bff}{\rm}
\begin{document}
\title{Emerging non-equilibrium bound state in spin current-local spin scattering}
\author{Fatih Do\u gan$^1$, Lucian Covaci$^1$, Wonkee Kim$^{1,2}$, and Frank Marsiglio$^1$}
\address{$^1$Department of Physics, University of Alberta, Edmonton, Canada\\
$^2$Department of Physics, University of Houston, Houston, Texas}

\date{\today}
\begin{abstract}
Magnetization reversal is a well-studied problem with obvious applicability in computer hard-drives.
One can accomplish a magnetization reversal in at least one of two ways: application of a magnetic field,
or through a spin current. The latter is more amenable to a fully quantum mechanical analysis. We
formulate and solve the problem whereby a spin current interacts with a ferromagnetic Heisenberg spin chain, to eventually reverse the magnetization of the chain. {\bff Spin-flips are accomplished through both elastic and inelastic scattering. A consequence of the inelastic scattering channel, when it is no longer energetically possible, is the } occurrence of a new entity: a non-equilibrium bound state (NEBS), which is an emergent property of the coupled local plus itinerant spin system. For certain definite parameter values the itinerant spin lingers near the local spins for some time, before eventually leaking out as an outwardly diffusing state. {\bff This phenomenon results in novel spin-flip dynamics and filtering properties for this type of system.}
\end{abstract}
\maketitle

\section{Introduction}

Most current computer hard drives utilize a technology for memory storage
which requires a switching of states involving magnetized spin. This switching
is accomplished through the application of magnetic fields in appropriate
directions. A theoretical understanding of this process is attained reasonably well
through a classical description via the Landau-Lifshitz-Gilbert equations \cite{review1,review2}.
These equations constitute a phenomenological description, since the required damping, whose
analytical form is even under some debate \cite{debate}, has various possible origins.

Just over a decade ago, however, theoretical proposals were made to
accomplish magnetization switching through spin transfer from {\em applied} spin currents
to magnetized spins \cite{slonczewski,berger}. A semi-classical description was used: the spin current was described by a plane wave, while the magnetized thin film that was to be flipped was described through
a classical magnetization vector. This problem became known as the `spin-torque' problem; the incoming
spin current exerts a torque on the local magnetization. It is noteworthy that in this problem a
phenomenological damping mechanism is not required to torque the magnetization in the direction of
the incoming spin current --- whereas the use of a magnetic field leads only to precession unless some
damping mechanism is introduced. The experimental observation of the `spin-torque' effect has met with some
limited success \cite{myers,tsoi,wegrowe,kiselev}.

Recently, a direct measurement of the spin torque vector depending on the
voltage has been made \cite{sankey}. Furthermore, the results of this experiment
imply that {\em inelastic} tunneling has an important effect on the spin transfer torque.
In fact, it appears that inelastic processes in the spin-flip scattering are
inherent \cite{balashov} for ferromagnetic systems.
In order to realize practical applications of the spin torque phenomenon,
it is important to reduce the critical current required to reverse the
magnetization of ferromagnets. A couple of experiments \cite{mangin,liu} have
demonstrated experimental methodologies to decrease the critical current.
As another signature of spin transfer, spin-torque induced magnetic vortex
phenomena are also observed \cite{pribiag,finocchio,strachan}.

The semi-classical picture seems to work well in a practical sense
\cite{bazaliy98,kim04,brataas03,kim06_cjp,jmmm}.
However, especially from a theoretical point of view,
some aspects are missing. Ultimately, spin transfer
is a quantum mechanical scattering problem, generally inelastic,
and so one would like to understand
the spin transfer process in terms of excitations of the ferromagnet.
Moreover, recent experimental work \cite{wang} has focussed on the
impact of a spin current on cobalt nanoparticles with diameter less than
$5nm$, which can be used to examine the spin torque exerted on
isolated nanoparticles. It has also been shown that it is experimentally feasible to
manufacture magnetic nanostructures (chains of 2-10 coupled atoms) \cite{hirjibehedin06}.
In this case, only a fully quantum mechanical description will suffice
because the quantum nature of the spin operator representing the stationary spins
in the nanoparticle is significant.

\begin{figure}
\begin{center}
\epsfig{figure=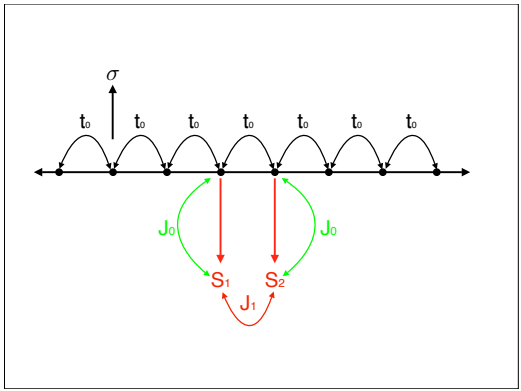,height=3.in,angle=90}
\caption{(Color online) A schematic of a lattice, on which an itinerant spin can hop (with hopping parameter $t_0$); it can interact with two stationary spins (indicated by downward pointing (red) arrows) with coupling strength $J_0$. The two stationary spins can interact with one another, with coupling strength $J_1$.}
\label{fig1}
\end{center}
\end{figure}

The scenario of an incoming (electron) spin, often modeled as a wave packet, whose spin degree of freedom
is coupled with local spins, has been advanced by a number of
workers \cite{avishai01,kim05,kim07,dogan08,ciccarelloPRL,ciccarello08}.
The coupling between the incoming spin
and the local spins is Kondo-like, while the local spins are themselves ferromagnetically coupled via
a Heisenberg exchange interaction. The model Hamiltonian is
\be%
H=-t_0 \sum_{<i,j>\sigma}c^{\dagger}_{i\sigma}c_{j\sigma}
-2\sum_{\ell=1}^{N_s}J_{0}{\bf\sigma}_{\ell}\cdot{\bf S}_{\ell}
-2\sum_{\ell=1}^{N_s-1}J_{1}{\bf S}_{\ell}\cdot{\bf S}_{\ell+1}%
\label{model_H}
\ee
where $c^{\dagger}_{i\sigma}$ creates an electron with spin $\sigma$
at site $i$, ${\bf S}_{\ell}$ is a localized spin operator at site $\ell$,
and $t_0$ is the hopping amplitude between nearest neighbor sites. The first term allows
an electron (of either spin) to propagate in a band that covers all space (here in one dimension),
while the second term is responsible for the Kondo-like interaction between the electron and the
local spins, with coupling constant $J_0$. This takes place over a finite chain of length $N_s$.
Finally, the last term models the Heisenberg exchange interaction with strength $J_1$ between the
local spins. For a ferromagnetic chain, $J_1 > 0$. Note, moreover, that if so desired, both $J_0$
and $J_1$ can depend on the position of the local spin within the finite chain. Fig. \ref{fig1} shows a schematic of this model.

{\bff The use of a wave packet to describe the incoming spin degree of freedom, and the subsequent `real-time' analysis of the scattering process allows us to examine the entire scattering process with very fine spatial and temporal resolution. While the present-day experimental capabilities do not quite match this fine resolution, we anticipate that probing on the time and length scales we use will be accessible
in the near future. In particular, in this work we identify an new feature which he have identified as a `non-equilibrium bound state' (NEBS), whose characteristics would require careful experimental identification. This phenomenon results because of an inelastic scattering process that is suppressed due to energy conservation. While an analytical approach does reveal some of the properties of a NEBS, the numerical wave packet calculations really allow us to see the non-equilibrium aspect of this phenomenon. Both calculations are presented here.

This paper is organized as follows. In the following section we outline means by which we solve the time-dependent problem. Some of our earlier work \cite{kim05,kim06,dogan08} used straightforward expansions in the basis states spanning the product Hilbert space of electrons moving on a lattice and stationary spins confined to a small portion of that same lattice. The present work uses a different method; the exponentiated Hamiltonian operator is expanded in a series utilizing Chebyshev polynomials \cite{weise08}.
This allows us to easily generate large scale numerical results, as described in the subsequent section. We formulate the problem for an arbitrary number of stationary spins (in principle, representing a magnetized thin film, whose magnetization is being flipped), but focus on two interacting stationary spins. This allows us to focus on the characteristic features of the larger system, including the NEBS, without the considerable complexity generated by the many scattering channels present when more than two stationary spins are used. Snapshots of the propagating wave packet reveal that in a particular region of parameter
space part of the wave packet `lingers' near the stationary spins. This feature is a signature of the NEBS.

In the fourth section we develop an analytical approximation to describe the same scattering process in the continuum limit. A preliminary decomposition of the problem, into less familiar but more useful basis states, first allows us to readdress the numerical results of Section III. This analysis identifies the NEBS with the position-dependent amplitude of one of these basis states. We further develop the analytical approximation to derive this amplitude, along with expectation values for the amount of spin-flip expected. Thus, while we lose the transparency of the time dependent (i.e. non-equilibrium) aspect of the problem, we clarify some of the physics of the bound state part. In Section V we conclude with some discussion concerning experimental observation of this NEBS.}

\section{theory}

We adopt the most straightforward approach to the scattering problem, and study the time evolution
of a wave packet, defined, at $t = 0$, as
\be
 \varphi(x)=\frac{1}{\sqrt{2\pi a^2}}e^{ik(x-x_0)}e^{\frac{-(x-x_0)^2}{2a^2}}
\ee
The calculation can take several routes at this stage. Consistent with the tight-binding formulation,
Eq. (\ref{model_H}), one can define a Hilbert space (with either open or periodic boundary conditions
left of the wave packet and far to the right of the local spins), with typically hundred's of lattice
sites on which the itinerant spin (hereafter referred to as the electron, or electron spin) can hop (see Fig.~1 for a schematic). One
can diagonalize Eq. (\ref{model_H}) on this Hilbert space and find the complete spectrum of eigenstates and eigenvalues with which one can construct the time evolution of the wave packet \cite{kim06,kim07,dogan08}.
However, we find that the parameter regime and maximum possible size of the local spin chain, for example, is severely restricted by computational expense within this approach.

Instead we choose to solve the time dependence directly, using the formal solution
\be
\Psi(x,t)=e^{-i\hat{H}t}\varphi(x).
\ee
A practical implementation of this solution is through the series expansion
\be
e^{-i\hat{H}t}=\sum_na_n\hat{Y}_n\ee
where $a_n$ are the coefficients of a complete orthonormal set of functions denoted by $Y_n$. A very useful
basis is provided by the Chebyshev polynomials, $T_n(x) \equiv {\rm cos}(n\cos^{-1}{x})$, with $T_0(X)=1, T_1(X)=X$ and $T_n(X)=2XT_{n-1}(X)-T_{n-2}(X)$ \cite{weise08}. For this expansion to be useful, the argument $X$ {\bff (here, a matrix)} is required to have norm less than unity, so a scaled version of the Hamiltonian is required (accompanied by a scaled
time variable):
\be
e^{-i \hat{H} t} = e^{-i\frac{\hat{H}}{\delta}\delta t}
=\sum_{n=-\infty}^{\infty}a_n(\delta t)T_n(-\frac{\hat{H}}{\delta})
=\sum_{n=-\infty}^{\infty}a_n(y)T_n(x)
\ee
where $y=\delta t$, and $x=-\frac{\hat{H}}{\delta}$.

There are two reasons for choosing this particular basis. First, the coefficients $a_n(y)$ can be written simply as \cite{cheb1}
\be
a_n(y)=\frac{1}{\pi}\int_{-1}^1\frac{dx}{\sqrt{1-x^2}}T_n(x)e^{(ixy)}=i^i{|n|}J_{|n|}(y),
\ee
where the $J_n(y)$ are Bessel functions of the first kind. Second, these polynomials have a recursion relation that allows us to use a more compact calculation of the expansion of the exponential of the Hamiltonian,
\be
T_{n+m} (x) = 2T_n(x)T_m(x)-T_{|n-m|} (x).
\label{nm}\ee
Using this equation we can rewrite the expansion up to a given order, $N^2$ as \cite{liang04}:
\bwt
\be
e^{i\frac{\hat{H}}{\delta}\delta t}\cong\sum_0^{N^2}a_iT_i=
\sum_0^Nb^0_iT_i+T_N[\sum_1^Nb^1_iT_i+...+T_N(\sum_1^Nb^k_iT_i+...+T_N\sum_1^{N}b^{N-1}_iT_i)...]
\ee
with
\be
b_i^k=\sum_{j=0}^{N-k}(mod(j,2)*A(j+k,k)a_{((j+k+1)*N-i)}+mod(j+1,2)*A(j+k,k)a_{((j+k)*N+i)})\ee
and the matrix elements $A(i,j)$ are defined by
\bea
A(i,j) = \left\{ \begin{array}{cc}
A(i-1,j)+2*A(i-1,j-1) & mod(i-j,2)=0\\
-A(i-1,j) & mod(i-j,2)=1\\
0 & i<j\\
\end{array}
\right.
\eea
with $A(0,0)=1$.
\ewt
This formulation allows for an efficient evaluation of the time evolution of the wave function,
such that large lattices can be studied, both for the electron spin, and for the stationary spin chain.

\section{Numerical Results}

\subsection{(a) non-interacting stationary spins}

\begin{figure}
\begin{center}
\epsfig{figure=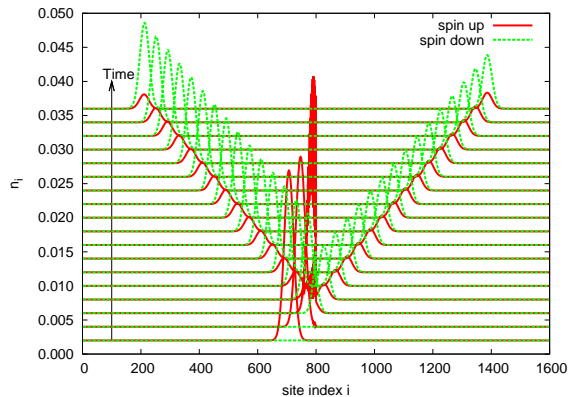,height=3.in,angle=-90}
\caption{(Color online) Time evolution of an electron wave packet, interacting with two local spins (located at sites 800 and 801). For the electron spin we use a tight-binding model with nearest neighbour hopping only; for reasons discussed in the text we use $k = \pi/2$. For this figure the coupling with local spins is given by $J_0=2.0t_0$, and the coupling between local spins is set to zero ($J_1=0$). The choice $J_1 = 0$ causes the time evolution of the electron spin to closely resemble the one with a single local spin previously reported in Ref. \onlinecite{kim05}. Subsequent time slices are displaced vertically for clarity.}
\label{figTime_ev}
\end{center}
\end{figure}

The result of a typical calculation is illustrated in Fig. \ref{figTime_ev}. Here, we have used 1600
lattice sites, and, at $t=0$ we have `launched' a wave packet centred around site 700 with a width
given by $a = 30$. The unit of length is the lattice spacing, which we take to be unity for convenience.
In all our figures we also take $t_0 \equiv 1$ as our energy scale.
All our results will utilize an {\bff initial} electron wave vector $k = \pi/2$, so that no wave packet broadening occurs \cite{kim06}. The incoming electron spin has $S = 1/2$, and, in the calculations in this paper, the stationary spins have $S = 1/2$. A series of snapshots is shown as time progresses forward. Initially only the incoming electron is present, represented as a Gaussian wave packet with only a spin-up component (shown as a solid (red) curve for the first time slice at the bottom). The initial conditions are such that all stationary spins (not shown but situated at sites 800 and 801) have $S_z = -1/2$ and the incoming electron spin has $S_z = 1/2$. As time advances the electron spin interacts with the stationary spins and
scatters. If there was only one stationary spin, the scattering would lead to 4 possibilities for the electron wave packet \cite{kim05}: it can either be reflected or transmitted, with either spin-up or spin-down. With two (or more) interacting stationary spins, inelastic scattering is also possible.
The choice of parameters in Fig. \ref{figTime_ev} is such that the result is similar to that expected from a single spin ($J_1 = 0$ here); after interacting with the local spins the wave packet both reflects and transmits with both spin components. The scattering is elastic which means the associated wave vectors are $\pm \pi/2$, so that no spreading of the wave packet occurs as time progresses (there is some intrinsic spread because two neighbouring scattering sites are involved).

The `final state' of both the electron and the local spins is readily defined by waiting for a period
of time after which the various electron components have separated a reasonable distance from the local spins. This is clear from the figure (the latest times shown clearly fulfil the above requirement) but
we will encounter special parameter regimes where this definition is not so clear, to be discussed
later.

\begin{figure}
\begin{center}
\epsfig{figure=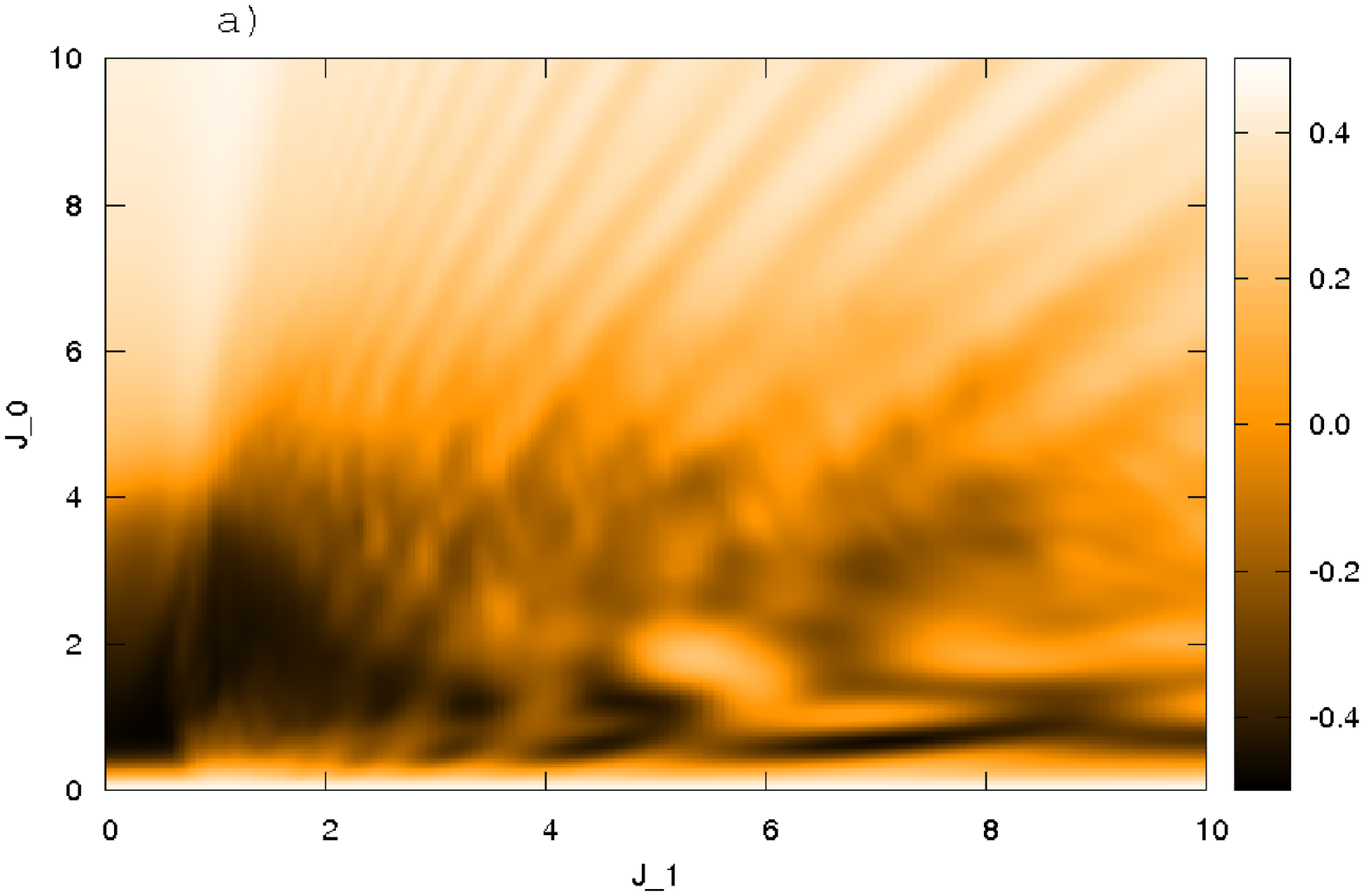,height=3.in,angle=-90}
\epsfig{figure=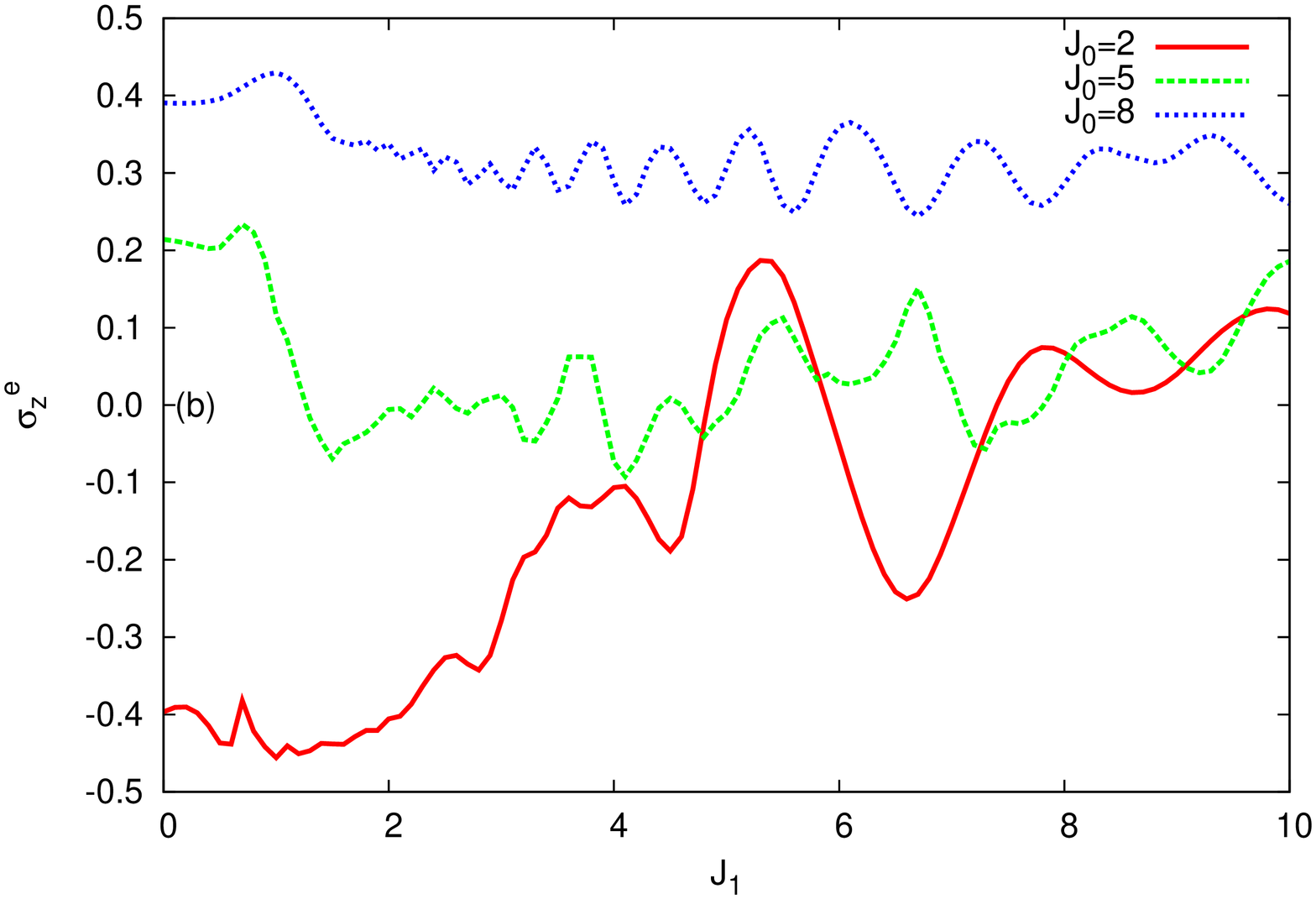,height=2.5in,angle=-90}
\caption{(Color online) (a) The $z$-component of the electron spin long after the electron wave packet has interacted with the local spins, as a function of both electron-spin coupling $J_0$ and spin-spin interaction $J_1$ for 20 local spins. The outcome is sufficiently complicated that we will focus on the problem with only two local interacting spins hereafter. (b) Slices are plotted as a function of $J_1$ for various values of $J_0$. As shown considerable complexity exists even in these plots.}
\label{fig20sites}
\end{center}
\end{figure}

\subsection{(b) `N' interacting stationary spins}

At the outset we wanted to understand how a (macroscopically) long spin chain interacts with an incoming electron spin to understand the effect of a spin current on a magnetic layer. With the technology discussed in the previous section for treating the time evolution of a coupled electron spin/local spins system, the study of reasonably long spin chains is indeed possible. However, the impact on the spin chain is sufficiently complex that this program was deemed overly ambitious for the present, even if we simply examine the impact on the electron spin as it emerges from the spin chain. Looking at `long times' after the interaction, the complexity in a series of figures like that in Fig. \ref{figTime_ev} for various values of $J_0$ and $J_1$ is enormous. The summary of such a plot is shown in Fig. \ref{fig20sites}, where the value of the $z$-component of the electron spin is shown after interaction with a spin chain consisting of 20 coupled $S=1/2$ spins. As a function of the interaction parameters $J_0$ and $J_1$ there are quite a number of visible ripple-like structures which no doubt are
related to the excitations that are populated through the inelastic scattering channels. This interpretation is reinforced by the observation that, for smaller spin chains, the number of ripples is reduced, as the number of possible internal excitations is reduced. Slices for fixed values of $J_0$ are illustrated in Fig. \ref{fig20sites}b, and again it is difficult to interpret all the various ripples. For this reason we focus, in the rest of this paper, on the simpler system where there are only two coupled local spins.

\begin{figure}
\begin{center}
\epsfig{figure=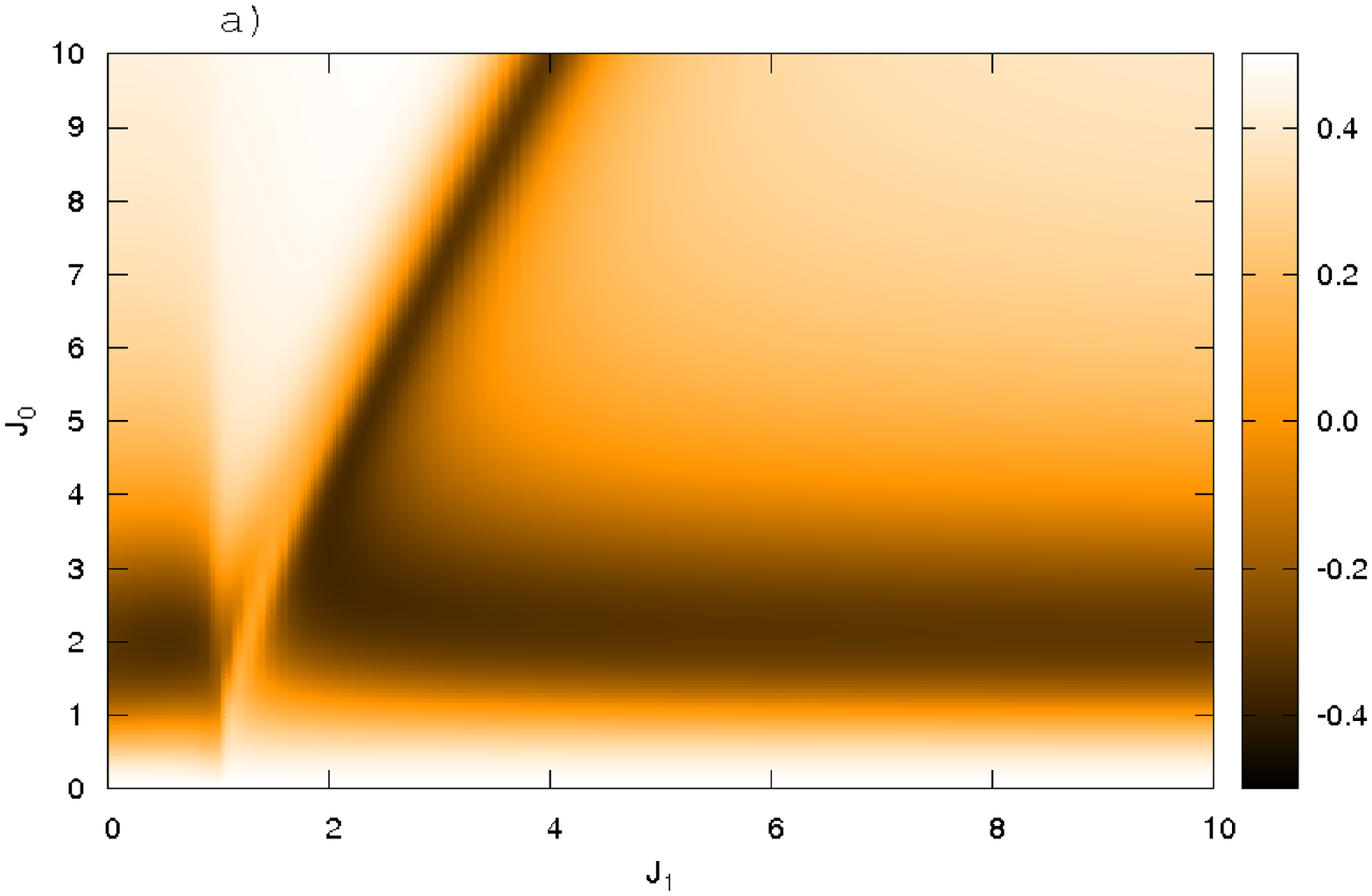,height=3.in,angle=-90}
\epsfig{figure=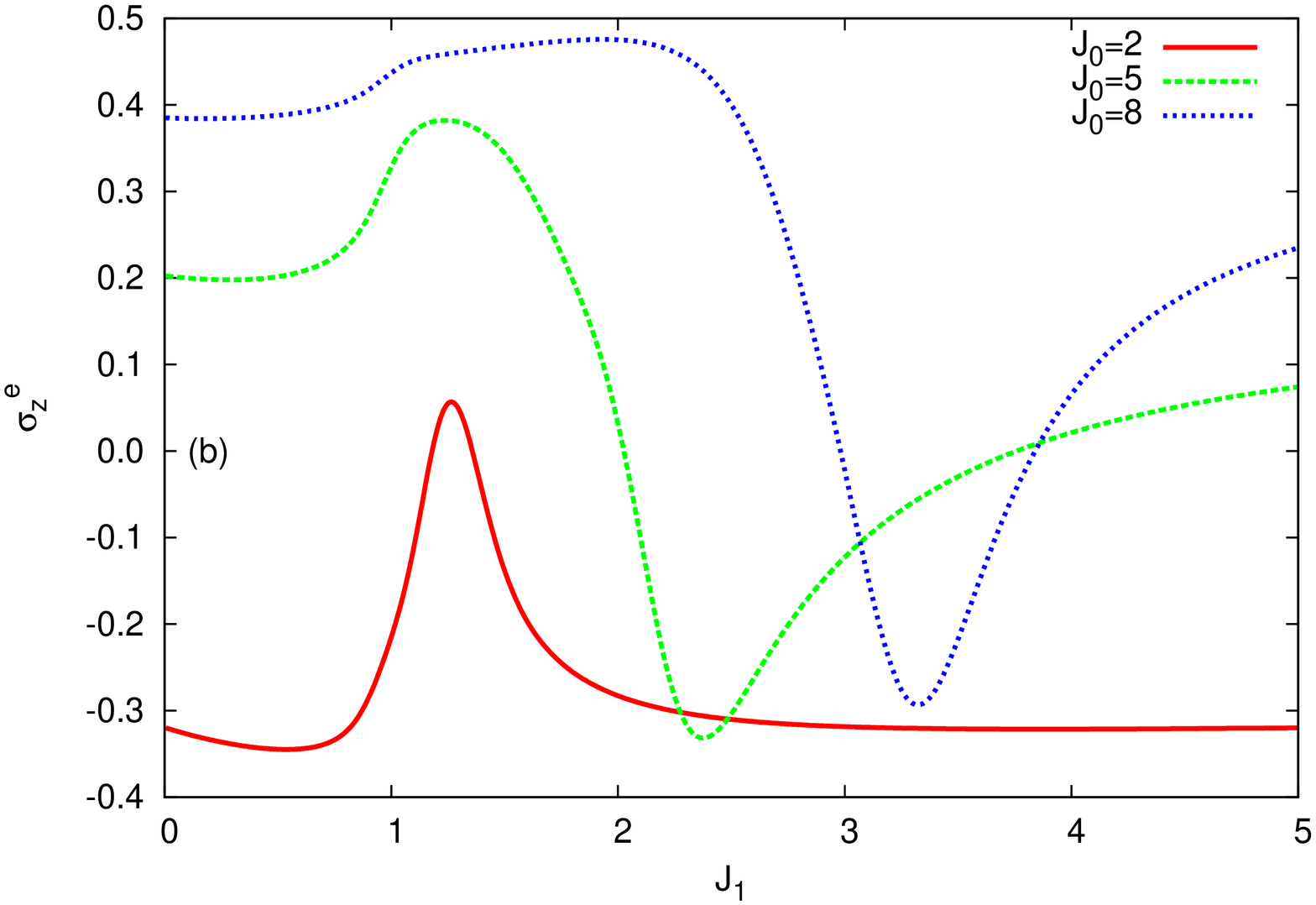,height=2.5in,angle=-90}
\caption{(Color online) (a) As in Fig. \ref{fig20sites}, the z-component of the electron spin long after the wave packet has interacted with the local spins, as a function of both electron-spin coupling $J_0$ and spin-spin interaction $J_1$ for 2 local spins. This plot is discussed extensively in the text. Note the horizontal band of strong spin-flip (dark) centred around $J_0=2 t_0$, broken only near $J_1 \approx 1.0t_0$. Smaller $J_1$ values result in independent behaviour by the 2 localized spins, while larger values of $J_1$ result in strongly coupled behaviour by the 2 local spins. A prominent but very slight change occurs along the vertical line at $J_1 = 1 t_0$, and a very obvious trough (i.e. a valley as far as the z-component of the electron spin is concerned) of spin-flip occurs as shown (in dark color) sloping up towards the right and exiting the graph at $(J_1,J_0) \approx (4t_0,10t_0)$. (b) Slices are plotted as a function of $J_1$ for various values of $J_0$. For $J_0 = 5t_0, 8t_0$ there is a definite valley corresponding to the dark `trough' just mentioned in the first plot, while, for $J_0 = 2t_0$, the behaviour is more complicated.}
\label{fig2sites}
\end{center}
\end{figure}

\subsection{(c) Two interacting stationary spins: inelastic scattering}

\begin{figure}
\begin{center}
\epsfig{figure=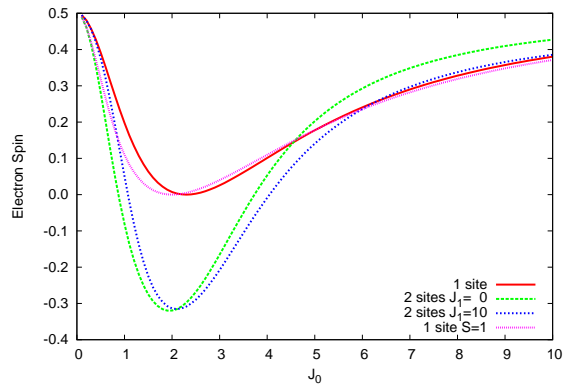,height=3.in,angle=-90}
\caption{(Color online) The $z$-component of the electron spin long after it has interacted with the local spin system, as a function of the Kondo coupling $J_0$. The solid (red) curve is the result for a single local spin with $S = 1/2$ \protect\cite{kim05}. Note that the maximum spin-flip occurs at an intermediate value of $J_0 \approx 2.3t_0$ \protect\cite{kim05}; when two local spins are present the result is similar, whether they are non-interacting ($J_1 = 0$) or strongly interacting ($J_1 = 10t_0$). As one would expect the degree to which the incoming electron can reverse its spin is much higher when interacting with more than one local spin.}
\label{figJ1_0}
\end{center}
\end{figure}

We first examine the long time behaviour of the electron spin. Fig. \ref{fig2sites} illustrates (in a color plot) the z-component of the electron spin once it has essentially left the vicinity of the two local spins, as a function of the Kondo coupling between electron spin and local spin, $J_0$, and the coupling between local spins, $J_1$. Curves are shown for the same quantity in Fig. \ref{fig2sites}(b), for specific values of $J_0$, as shown; these correspond to horizontal sweeps across the first plot. In Fig. \ref{figJ1_0} vertical sweeps across the first plot in Fig. \ref{fig2sites} are shown, along with the result for a single local spin \cite{kim05}. {\bff The sweeps are plotted for extreme values of $J_1$, and
avoid the complicated region characterized by a `trough' (colored dark) of significant spin flip rising upwards to the right, and leaving the plot area at $(J_1,J_0) \approx (4,10)t_0$. This `trough' region will be discussed in detail in the next subsection.}

A considerable amount of information is contained in Fig. \ref{fig2sites}. The horizontal band of strong spin-flip (dark) centred around $J_0=2 t_0$ is further illustrated for specific values of $J_1$ in Fig. \ref{figJ1_0}, as a function of $J_0$ (the dark horizontal band in Fig. \ref{fig2sites}a corresponds to the minima visible in Fig. \ref{figJ1_0}). Whether or not the local spins are strongly coupled, the net effect on the electron spin is similar, and in qualitative agreement with what happens when only a single localized spin is present \cite{kim05} (solid (red) curve in Fig. \ref{figJ1_0}.) As already described for a single local spin \cite{kim05,kim06,kim07,dogan08}, the maximum spin flip occurs near $J_0 = 2t_0$; for very small values or very large values of $J_0$ the impact on the electron spin goes to zero. 

\begin{figure}
\begin{center}
\epsfig{figure=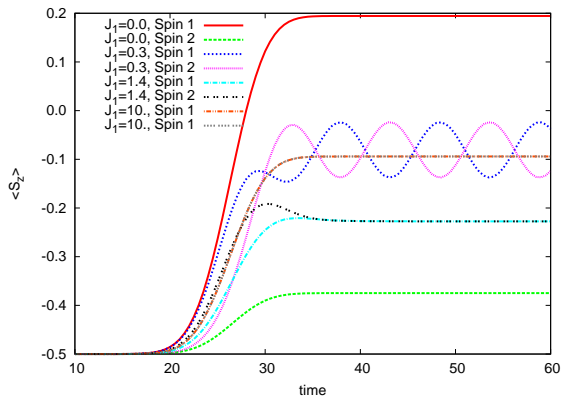,height=3.in,angle=-90}
\caption{(Color online) The $z$-component of the two local spins for different spin-spin interaction strengths $J_1$, all  for $J_0=2t_0$. For $J_1=0$ the two spins are essentially independent of one another, while for $J_1=10t_0$ the two local spins are locked together with the same value as a function of time.}
\label{figSz}
\end{center}
\end{figure}

The reaction of the local spins {\em does} depend on the value of the coupling between local spins, as illustrated in Fig. \ref{figSz}, where the $z$-component of the two local spins are shown as a function of time for various values of $J_1$.
For zero coupling they react independently (except the second local spin `sees' only part of the incoming electron spin, because it has already scattered and spin-flipped off the first), while for low coupling some precession occurs. At high values of the coupling, the two local spins are essentially locked together.

Referring again to Fig. \ref{fig2sites}, a subtle change occurs as $J_1$ passes through $t_0$ for all values of $J_0 > t_0$; this is
more clearly seen in Fig. \ref{fig2sites}(b), where a small rise occurs in the z-component of the electron spin as $J_1/t_0$ crosses unity. For $J_0 = 2t_0$ the increase is considerable, followed by a peak and then a monotonically decaying result. This is in contrast to the other two curves which also show a minimum. In fact these two curves are more `generic'; inspection of Fig. \ref{fig2sites}(a) shows that $J_0 = 2t_0$ passes right through the middle of the dark band which was discussed above. This region of the $J_0 - J_1$ phase diagram is fairly complicated --- the three energy scales are all similar in size and no simple picture emerges.

Focusing on the larger values of $J_0$, the small increase in the $z$-component of the electron spin shown in Fig. \ref{fig2sites}(b) (also
visible in Fig. \ref{fig2sites}(a) as a faint but abrupt break along the vertical line $J_1 = t_0$) can be understood as follows. First note that this increase signals a {\em decrease} in the spin-flip interaction. Recall that the electron spin is propagated with wave vector $k = \pi/2$. This means that its kinetic energy is effectively $2t_0$ --- the dispersion relation $\epsilon(k) = -2t_0 \cos{(ka)}$ gives $\epsilon(k = \pi/2) = 0$, but $2t_0$ is the energy {\em with respect to the bottom of the band}. Thus, the electron has a maximum energy $2t_0$ that can be deposited into the local spin system through the Kondo-like coupling $J_0$. On the other hand, for a two spin system there is only one non-zero excitation energy --- it is $E_{ex} = 2J_1$ --- and this is essentially the spin wave energy for a two spin system, as can be readily
ascertained from the solution to the problem of two ferromagnetically coupled Heisenberg spins \cite{ashcroft76}. For $J_1 > t_0$ this mode of inelastic scattering is no longer possible, so the amount of spin-flip scattering decreases, as indicated in the figures.

\begin{figure}
\begin{center}
\epsfig{figure=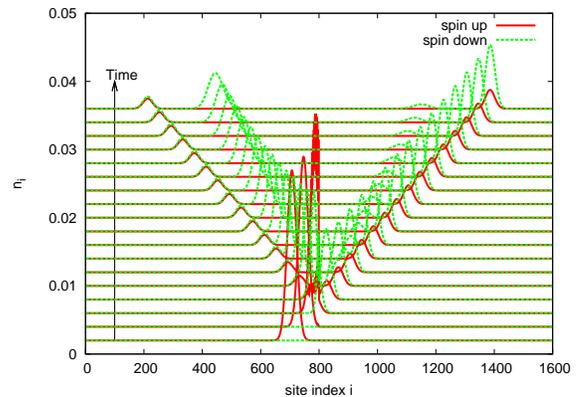,height=3.in,angle=-90}
\caption{(Color online) A series of snapshots of the electron wave packet, with both spin up (solid, red curves) and spin down (dashed, green curves). Note that spin down components are scattered both elastically (same speed as incoming wave packet) and inelastically (slower speed, indicated by a more vertical profile on this plot). The scattering occurs off of two local spins, located at sites $800$ and $801$, ferromagnetically coupled with $J_1/t_0 = 0.8$; we used $J_0/t_0=2.0$.}
\label{figTime_ev2}
\end{center}
\end{figure}

An explicit demonstration of this mode of scattering is provided in Fig. \ref{figTime_ev2}, where a series of snapshots of the electron wave packet is shown as a function of position. In contrast to Fig. \ref{figTime_ev} a second set of peaks is evident, all in the spin-flip channel (i.e. $z$ component of electron spin is $-1/2$) moving more slowly (hence inelastic scattering) both to the left (reflection) and to the right (transmission). As $J_1 \rightarrow t_0$ the speed of this wave packet approaches zero (so the extra wave packets will appear almost vertically in a plot like Fig. \ref{figTime_ev2}). For more and more coupled local spins many more inelastic channels are available for scattering, which in part explains the complexity in Fig. \ref{fig20sites}.

\subsection{(d) Two interacting stationary spins: the non-equilibrium bound state (NEBS)}

\begin{figure}
\begin{center}
\epsfig{figure=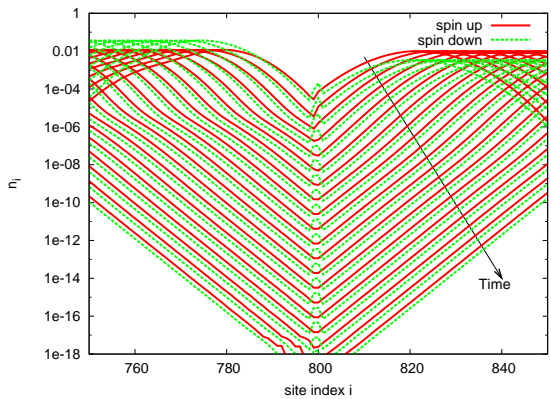,height=3.in,angle=-90}
\caption{(Color online) A series of snapshots of the electron wave packet, with both spin up (solid, red curves) and spin down (dashed, green curves). Note that time progresses forward as one moves from curve to curve {\em downwards}, and also note the logarithmic scale for the ordinate. By the last times shown the usual Gaussian wave packet peaks have disappeared off to the sides; what remains, however, is a small peak located near the local spins. We refer to this as a non-equilibrium bound state (NEBS); justification for this name will come in the next section. Note that this small peak exists only in the flipped spin channel. The scattering occurs off of two local spins, located at sites $800$ and $801$, ferromagnetically coupled with $J_1/t_0 = 3.1$ and with a Kondo-like coupling $J_0/t_0=8.0$; with reference to Fig. \protect\ref{fig2sites} these parameters place us in the middle of the dark colored trough of enhanced spin flip scattering.}
\label{figTime_ev4}
\end{center}
\end{figure}

The most striking feature in Fig. \ref{fig2sites} is the trough (dark color) that extends upwards to the right, and exits the graph at $(J_1,J_0) \approx (4,10)t_0$. This trough represents a domain in the coupling space in which the spin-flip interaction persists more than expected, and is roughly associated with a `resonance' behaviour. The evidence for this is very difficult to glean from the numerical calculations --- we will have more to say based on analytical work to be presented in the next section. Nonetheless, examination of the numerical results for a particular set of parameters on a  logarithmic scale shows an unusual feature, as illustrated in Fig. \ref{figTime_ev4}, for relatively high parameter values of electron-spin coupling, $(J_1,J_0) = (3.1,8)t_0$. On this scale the Gaussian wave packets are outside the displayed region at the latest times shown (note that time progresses as one moves {\em down} from curve to curve, opposite to the progression shown in previous plots). The feature in question is the rather small peak located at the local spin sites (near site $800$ and $801$) that persists, albeit with strongly diminishing amplitude, for all times shown. This peak forms only for the spin down component of the electron; its amplitude decays away
{\em in both spin channels}
presumably through a diffusive process, so eventually the electron has scattered entirely.
We refer to this state as a non-equilibrium bound state (NEBS); this name will be further justified in the next section.

\begin{figure}
\begin{center}
\epsfig{figure=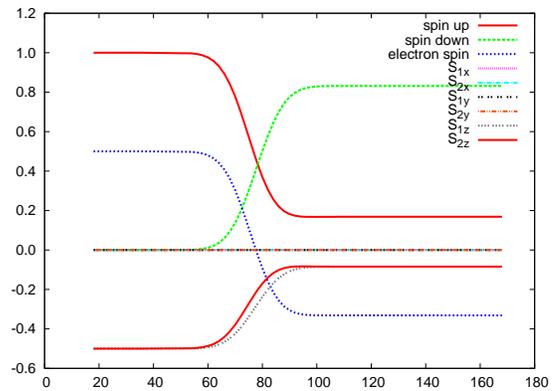,height=3.in,angle=-90}
\caption{(Color online) The $S_x$, $S_y$, and $S_z$ components of the two local spins for the parameter set discussed in the previous figure. Note that $S_x$ and $S_y$ remain equal to zero (due to the initial conditions, as explained in Ref. \protect\cite{kim05}), while the $S_z$ components change, although in reverse order than one would naively expect. This is an instance where the classical notion of a `spin vector' that rotates into the direction of the spin current while maintaining a constant magnitude is completely inapplicable.}
\label{figsxyz}
\end{center}
\end{figure}

In Fig. \ref{figsxyz} we show the various components of the local spins as a function of time, along with the electron spin. The $S_x$ and $S_y$ components remain fixed at zero (because of the initial conditions on these spins \cite{kim05}), while the $S_z$ components flip partially and remain at the same value long after the flipping process has terminated. In the intermediate stages, however, they are {\em not} locked together, and remarkably, the 2nd spin flips {\em before} the first. This reversal of the expected order of flipping occurs only for parameters in the `trough' region; otherwise the local spin first encountered by the incoming electron spin is the first to flip. While this phenomenon is clearly connected to the NEBS, we do not have a simple explanation for the spin flip reversal.

These results illustrate the variety of different behaviour possible for the spin flip scattering process as a function of $J_0$ and $J_1$. We now turn to an analytical approach to gain some further insight into the problem.

\section{Analytical Plane Wave Approximation}

\subsection{(a) a change of basis}

The problem of an incident spin represented as a plane wave scattering off of an impurity with a contact Kondo-like spin-spin interaction was solved analytically in Ref. \cite{kim05}. In that problem we made use of the initial conditions and conservation of angular momentum to simplify the problem. Here we do the same, and utilize initial conditions such that the $S_z$ component of the incoming electron spin is $+1/2$, while those of the two stationary spins are each $-1/2$.

The one-dimensional version, written in free space, has a Hamiltonian which can be written:
\be
H = -\frac{\hbar^2}{2m} \frac{d^2}{dx^2} - 2 J_0 [\hat{\sigma}
\cdot \hat{S}_1 \delta(x) + \hat{\sigma} \cdot \hat{S}_2\delta(x-a)] - 2 J_1
\hat{S}_1 \cdot \hat{S}_2.
\label{model_h_free}
\ee
The wave function for this problem consists of a spatial component which describes the electron spin amplitude, and a spin part which describes the spin state of the incoming electron and the two stationary spins (here located at positions $x=0$ and $x=a$). The Hilbert space concerning the spin degrees of freedom has an overall size of $2^3 = 8$ (for $S=1/2$ spins) However, utilizing the conservation of total $S_z$ reduces this number to $3$. As already stated, the initial state, in
Dirac notation, is  $|\uparrow\downarrow\downarrow\rangle $,  where the first arrow represents the $z$-component of the electron spin, and the next two arrows indicate the respective $z$-components of the two local spins. Once the electron spin interacts with the local spins, two more spin states are possible, $|\downarrow\uparrow\downarrow\rangle$ and  $|\downarrow\downarrow\uparrow\rangle$. {\bff In our numerical results, we followed two separate routes: in cases with the initial configuration as depicted here, we used this fact to reduce the Hilbert space to these three spin states, which sped up the calculations considerably. Alternatively, when the initial configuration was not so straightforward (and did not have a definite total $S_z$, for example), we used all 8 basis states.}

When we begin with an initial configuration like $|\uparrow\downarrow\downarrow\rangle $, we can 
combine these spin states into combinations with both good total $S_z$ {\em and} good total $S$ to give rise to the following basis set \cite{schiff55}:
\bea
|\psi_1 \rangle &=& \frac{1}{\sqrt{3}}
(|\downarrow\downarrow\uparrow\rangle +
|\downarrow\uparrow\downarrow\rangle +
|\uparrow\downarrow\downarrow\rangle)
\label{spinh}\\
|\psi_2 \rangle &=& \frac{1}{\sqrt{6}}
(|\downarrow\downarrow\uparrow\rangle +
|\downarrow\uparrow\downarrow\rangle -2
|\uparrow\downarrow\downarrow\rangle)
\label{spinf}\\
|\psi_3 \rangle &=& \frac{1}{\sqrt{2}}
(|\downarrow\downarrow\uparrow\rangle -
|\downarrow\uparrow\downarrow\rangle).
\label{sping}
\eea
\begin{figure}
\begin{center}
\epsfig{figure=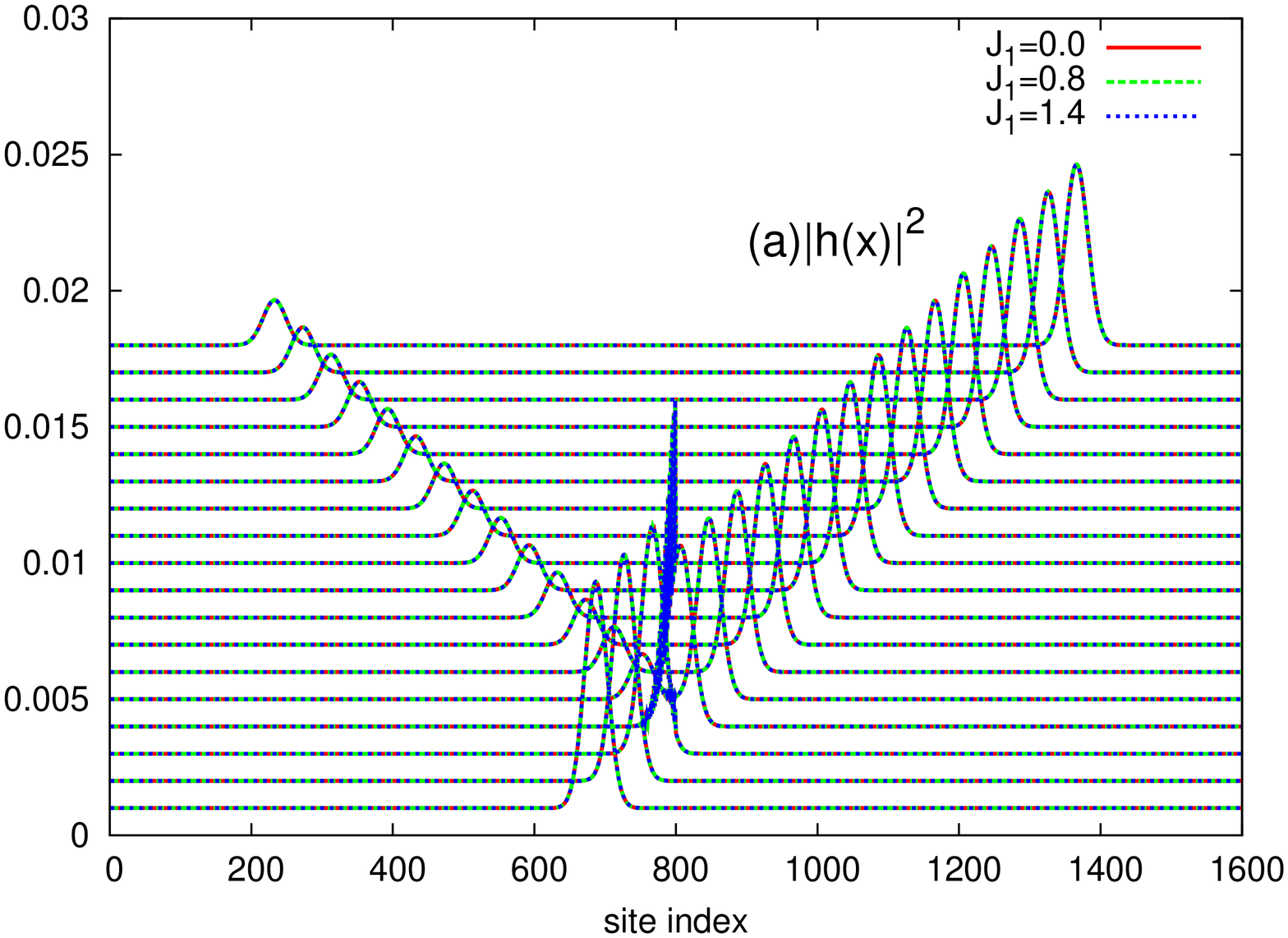,height=2.5in,angle=-90}
\epsfig{figure=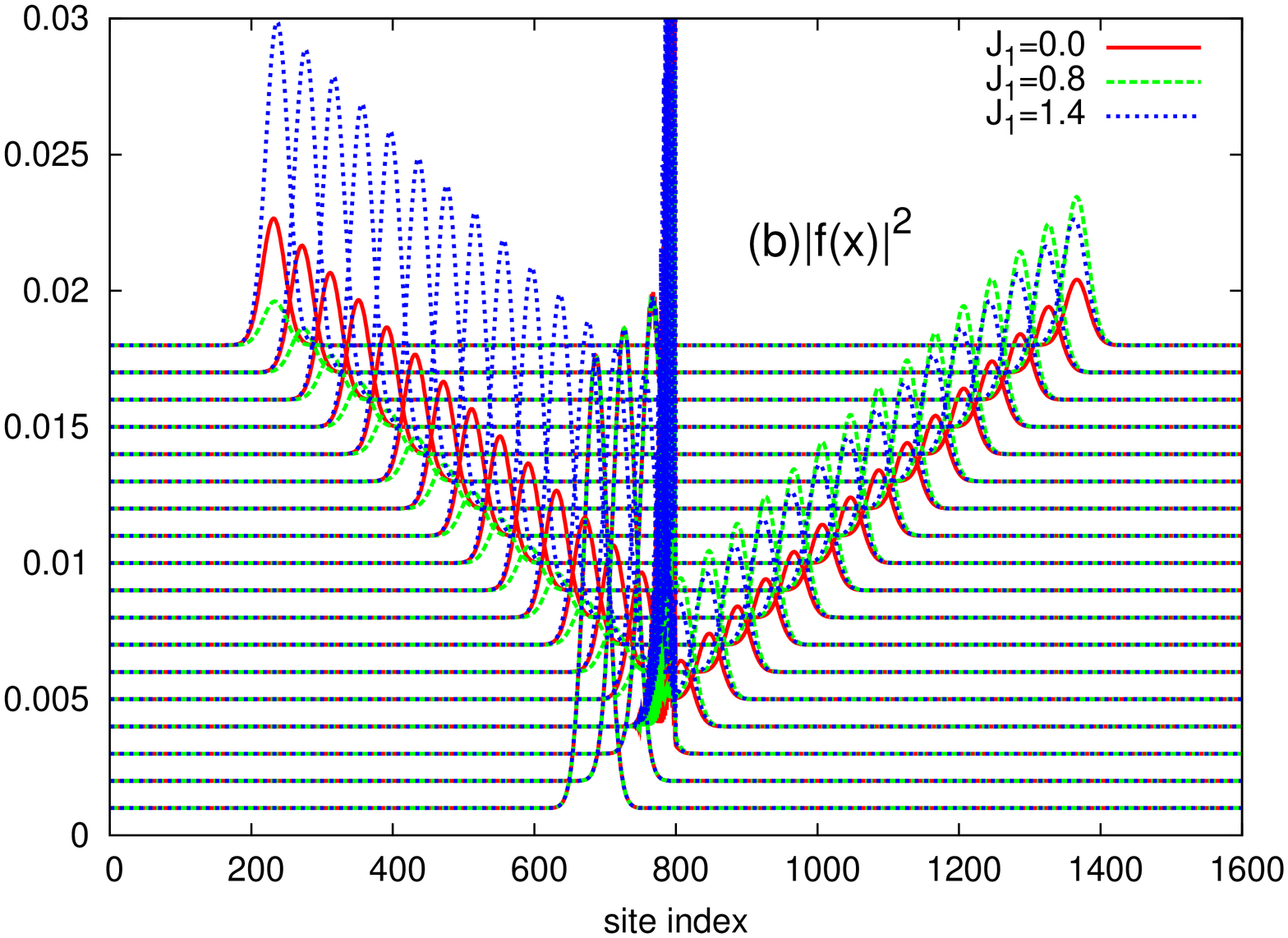,height=2.5in,angle=-90}
\epsfig{figure=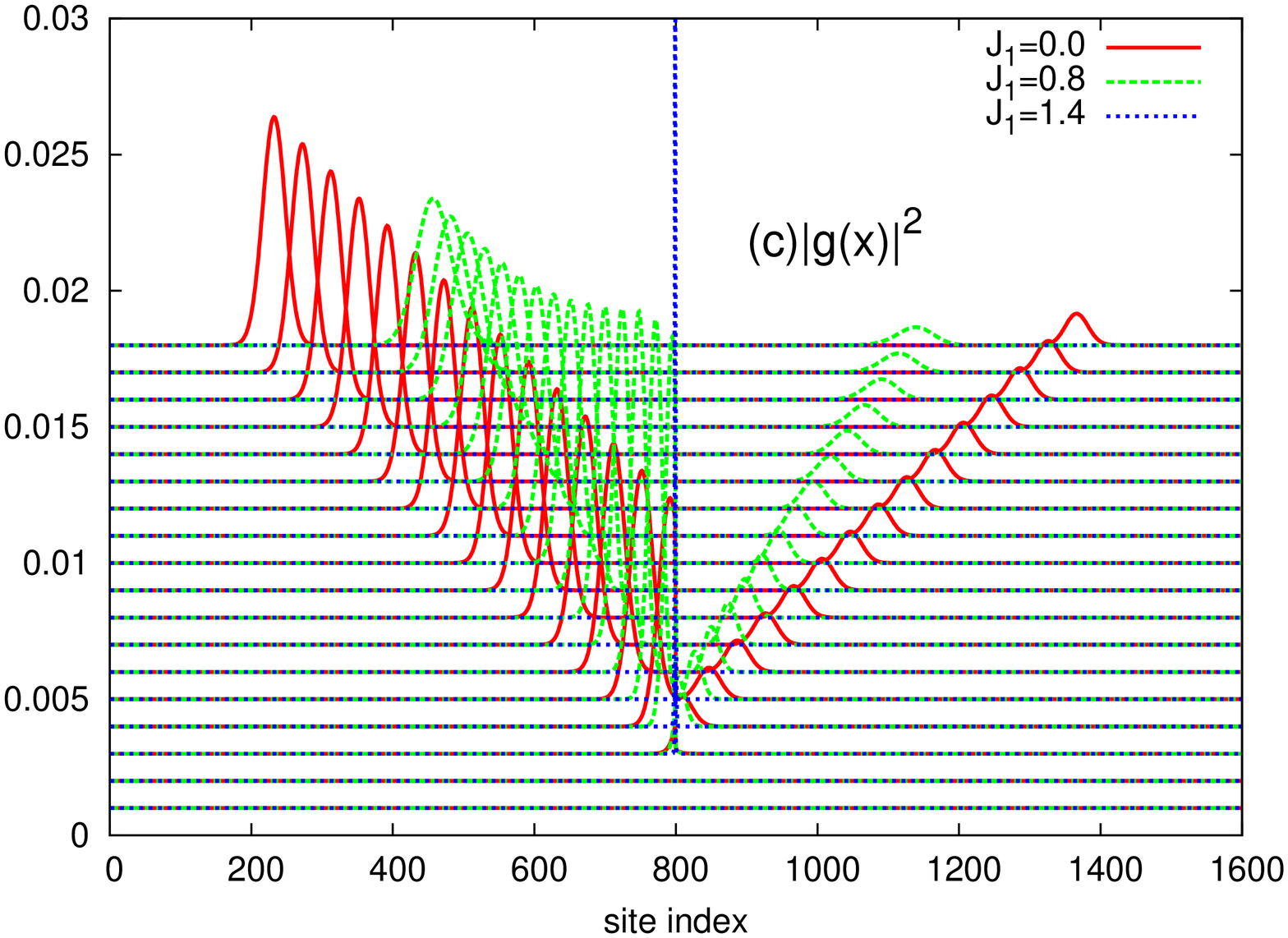,height=2.5in,angle=-90}
\caption{(Color online) The time evolution of the magnitudes (a) $|h(x)|^2$, (b) $|f(x)|^2$, and (c) $|g(x)|^2$, as defined by the basis set, Eqs. (\protect\ref{spinh}-\ref{sping}). These plots apply for $J_0=2t_0$ and the values of $J_1$ indicated. Note that in (a) the plots are identical for all 3 values of $J_1$, as motivated by the structure of Eq. (\ref{eq_h}). In (b) and (c) differences are apparent; note that in (c) no amplitude is present before the time of scattering, and, furthermore, as one enters the 'trough' region ($J_1 = 1.4t_0$) $|g(x)|^2$ consists of a single sharp peak near the local spins. In time this peak diffuses outwards, but there is no wave packet component. }
\label{fig_state_hfg}
\end{center}
\end{figure}
Writing the wave function as
\be
|\psi_(x) \rangle = h(x) |\psi_1 \rangle  +  f(x) |\psi_2 \rangle  + g(x) |\psi_3 \rangle,
\label{total_psi}
\ee
then appropriate projection on to the spin basis states
results in the three equations:
\bwt
\bea -\frac{\hbar^2}{2m}\frac{d^2h}{dx^2} - 2 J_0 \frac{\hbar^2}{4}\Big(\delta(x) + \delta(x-a)\Big)h &=& \epsilon_{el}h
\label{eq_h} \\
-\frac{\hbar^2}{2m} \frac{d^2f}{dx^2} + J_0 \frac{\hbar^2}{2} [\delta(x)(2f - \sqrt{3}g)+\delta (x-a)(2f+\sqrt{3}g)] &=& \epsilon_{el}f
\label{eq_f}\\
- \frac{\hbar^2}{2m}\frac{d^2g}{dx^2} - \sqrt{3} J_0 \frac{\hbar^2}{2} [\delta(x) - \delta(x-a)]f &=& (\epsilon_{el} - 2 J_1 \hbar^2) g
\label{eq_g}
\eea
\ewt
where $\epsilon_{el}$ is the kinetic energy of the incoming electron. Note that the first equation results from the $S_{\rm tot} = 3\hbar/2$ sector, and remains decoupled, while the second two are part of the $S_{\rm tot} = \hbar/2$ $S_{\rm tot z} = -\hbar/2$ doublet. The state with spatial wave function $g(x)$, governed primarily by the third equation, exists exclusively because of the possible inelastic scattering process. Eq. (\ref{sping}) indicates that it contains only the spin down component of the scattered electron, and, given our initial conditions, exists only after scattering. It is `fueled' through the $f(x)$ component, which, as Eq. (\ref{spinf}) indicates, contains a component corresponding to the incoming electron spin (with $\sigma_z = +\hbar/2$). That the $g(x)$ component corresponds to inelastic scattering is indicated by the eigenvalue on the right-hand-side of Eq. (\ref{eq_g}), with value $\epsilon_{el} - 2\hbar^2J_1$, which shows that an energy $2\hbar^2J_1$ is left behind in the form of a spin wave excitation in the local spin system, as explained in the previous section.  The first two equations, Eqs. (\ref{eq_h}) and (\ref{eq_f}), each have eigenvalue $\epsilon_{el}$, showing that the kinetic energy of the incoming electron is conserved (elastic scattering). Note that this still results in spin-flip scattering; it is just that the two local spins are scattered by the same amount, so that they remain in their coupled ground state.

\subsection{(b) a re-examination of the numerical solutions}

Eqs. (\ref{eq_h}-\ref{eq_g}) can be readily solved analytically, and we will come to that solution shortly. However, already Eqs. (\ref{spinh}-\ref{sping}) serve the important task of directing our attention to specific linear combinations of the spin states, as indicated. The numerical solutions presented in the previous section were classified only according to the $z$-component of the electron spin. We now essentially re-plot those results, as separate amplitudes $h(x)$, $f(x)$, and $g(x)$, in Fig. \ref{fig_state_hfg}. {\bff Note that Eqs. (\ref{eq_h}-\ref{eq_g}) were derived for the continuum model defined by Eq. (\ref{model_h_free}); nonetheless the role of the various amplitudes, described at the end of the previous subsection, applies equally well to the numerical results of the original tight-binding model.}

To demonstrate this, in Fig. \ref{fig_state_hfg}(a) we plot the magnitude $|h(x)|^2$ vs. position for a number of time slices, for three different values of $J_1$. As predicted by Eq. (\ref{eq_h}), there is no dependence on $J_1$. It is important to note that these results still represent numerical solutions to the tight-binding model presented in the previous section; while we could have used the spin components as listed in Eqs. (\ref{spinf}-\ref{sping}) as a basis set, these numerical solutions do not `use' the analytical structure of Eqs. (\ref{eq_h}-\ref{eq_g}). Hence only one set of curves is visible (for $J_1 = 1.4t_0$) as this set is identical to and covers entirely the sets corresponding to the other two values of $J_1$.
\begin{figure}
\begin{center}
\epsfig{figure=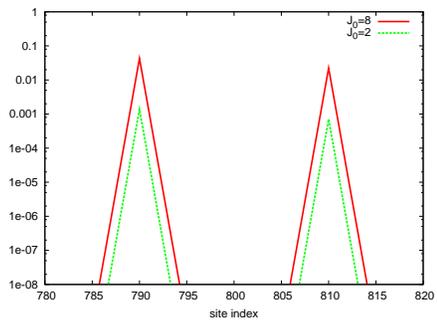,height=2.4in,angle=-90}
\caption{(Color online) A plot of $|g(x)|^2$ (see Eq. (\protect\ref{total_psi})) vs. $x$. This 'snapshot' is taken immediately following the initial scattering of the electron spin with the two local spins, situated at sites $790$ and $810$, i.e. they have been separated for clarity. We use $J_1 = 3.1t_0$, so it is clear that for parameter values that fall on the 'trough' ($J_0=8t_0$) the component of the wave function associated with inelastic scattering (i.e. $g(x)$) is significantly enhanced (almost two orders of magnitude) compared with the region away from the trough.}
\label{fig_g_20apart}
\end{center}
\end{figure}

In contrast, the other two components, plotted in Figs. \ref{fig_state_hfg} (b) and (c), are dependent on the value of $J_1$.
In both cases the amplitudes of transmitted and reflected wave packet depend quantitatively on the value of $J_1$. Note, moreover, that the amplitude $g(x)$ has no `incoming' wave packet. As explained earlier this amplitude is generated entirely by the scattering process. Also note that for $J_1/t_0 < 1$ (i.e. $J_1/t_0 = 0.8$ in Fig. \ref{fig_state_hfg}) the slow moving piece belongs entirely to $g(x)$ while the fast moving one belongs entirely to $f(x)$.

To see the role of the $g$ component of the state more clearly, we separate the two local spins by 20 sites, and project out the $g$ component from the numerical solution, using Eq. (\ref{total_psi}). In Fig. \ref{fig_g_20apart} 
we show on a log scale the magnitude of the $g$ component, $|g(x)|^2$ as a function of position; the two local spins are now located at sites $790$ and $810$. The parameters $(J_0/t_0,J_1/t_0)=(8,3.1)$ (solid curve) situate the regime on the 'trough' so apparent in Fig. \ref{fig2sites}, whereas $(J_0/t_0,J_1/t_0)=(2,3.1)$ (dashed curve) puts one well away from the trough. This snapshot is taken immediately after the initial scattering takes place, and it is clear that the $g$ component is almost 2 orders of magnitude larger on the trough (solid curve) than off (dashed curve). A similar result holds for large values of $J_0$.


\subsection{(c) analytical solution}

An analytical solution of the problem with plane waves through Eqs. (\ref{eq_h}-\ref{eq_g}) is possible, though tedious. One defines three regions in space, and defines the wave function in a piecewise continuous manner, as is done commonly in undergraduate physics texts.


With $k \equiv \sqrt{2m\epsilon_{\rm el}/\hbar^2}$, and $Q \equiv \sqrt{2m(2 \hbar^2 J_1 - \epsilon_{\rm el})/\hbar^2}$, the wave function can be written
\be
h(x) = \left\{ \begin{array}{cc}
h_I     e^{ikx} + u_I e^{-ikx}, \phantom{aaaaaaaa} x < 0  \\
h_{II}  e^{ikx} + u_{II} e^{-ikx}, \phantom{aaa} 0<x < a\\
h_{III} e^{ikx}, \phantom{aaaaaaaaaaaaaa} x > a \\
\end{array}
\right\}
\label{hcoeff},
\ee
\be
f(x) = \left\{ \begin{array}{cc}
f_I     e^{ikx} + r_I e^{-ikx}, \phantom{aaaaaaaa} x < 0   \\
f_{II}  e^{ikx} + r_{II} e^{-ikx}, \phantom{aaa} 0<x < a\\
f_{III} e^{ikx}, \phantom{aaaaaaaaaaaaaa} x > a \\
\end{array}
\right\}
\label{fcoeff},
\ee
and
\be
g(x) = \left\{ \begin{array}{cc}
g_I   e^{Qx}, &  \phantom{aaaaaa} x < 0\\
g_{II} e^{-Qx} + s_{II} e^{Qx},& \phantom{aaa} 0<x < a\\
g_{III}e^{-Qx}, &\phantom{aaaaaaa} x > a \\
\end{array}\right\}
\label{gcoeff}.
\ee
Four conditions relate the various coefficients defining $h(x)$ to the incoming amplitude $h_I$ in Eq. (\ref{hcoeff}); similarly eight conditions determine the $f$ and $g$ coefficients in terms of the incoming amplitude $f_I$. The results for $h(x)$ are standard and can be found in many undergraduate texts, while, for $f$ and $g$, the result is not standard, but is nonetheless straightforward. Also note that we have written the wave function for the more relevant condition $2 \hbar^2 J_1 > \epsilon_{\rm el}$, in which case the function $g(x)$ is exponentially decaying; the alternative $2 \hbar^2 J_1 < \epsilon_{\rm el}$
is straightforward and gives a propagating wave solution, with wave vector $q = \sqrt{2m(\epsilon_{\rm el} - 2 \hbar^2 J_1)/\hbar^2}$. This latter case corresponds to the situation whereby a spin wave excitation is energetically allowed, so that a spin-flipped wave packet will emerge from the stationary spins at a reduced speed, as we have already seen in the numerical solution in Fig. \ref{figTime_ev2}.

When inelastic scattering is not allowed by energy considerations, it is not clear what will happen.
Our intuition suggests that the stationary spins will respond as one, and so the spin-flip process will
resemble that expected for scattering from a single spin (which, as we demonstrated earlier, is not so different from scattering off decoupled stationary spins ($J_1 = 0$)). Inspection of Fig. \ref{fig2sites}
shows that this is indeed the case, {\em except for the trough region previously identified}. It is precisely in this regime that a peculiar enhancement of spin-flip scattering occurs, which we now argue is connected to the effective bound state (NEBS) defined by Eq. (\ref{gcoeff}).





The solutions can readily be written down by using the definitions,
$\alpha \equiv J_0/k$ and $\beta \equiv J_0/Q$ (the mass $m$ is set equal to unity), and the terms $v \equiv \alpha \biggl\{1 - {3\beta \over 4}(1 - e^{-Qa} e^{ika})\biggr\}$ and $u \equiv \alpha \biggl\{1 - {3\beta \over 4}(1 - e^{-Qa} e^{-ika})\biggr\}$ appear often. Note that for real $Q$ these are complex conjugates of one another. However, these expressions (and the ones immediately following) are valid for high electron kinetic energy as well, where $\epsilon_{\rm el} > 2J_1$, and so it follows that $Q= -iq$ with $q$ now real, and $u$ and $v$ are no longer complex conjugates of one another. We find, for example,
%
\be
{g_{II} \over f_{I}} = {\sqrt{3} \over 2} \beta { 1+iv-iue^{2ika} \over (1+iv)^2 + u^2e^{2ika}},
\label{g2}
\ee
with similar expressions for the other coefficients.
To see how effective the spin flip process is, we can calculate either the expectation value of the electron spin, $\langle \sigma_z \rangle$, or the spin torque, $N_{\rm zx}$ \cite{kim06_cjp}. For the two local spin system used here, the latter is given in terms of the former as $N_{\rm zx} = k(1/2 - \langle \sigma_z \rangle)$. As in the earlier numerical results, the quantity $\langle \sigma_z \rangle$ will remain near $0.5$ (the initial electron spin value) if very little spin flip occurs, whereas this quantity will deviate most from $0.5$ (and even become negative) when significant spin flip occurs. Note that with the plane wave solution given in Eqs. (\ref{hcoeff}-\ref{gcoeff}), the problem is no longer time dependent; one envisions a continual influx of current (this is $f_I$) while reflected and transmitted plane waves (of both spin type) take on 'steady-state' values \cite{confusion}.

The calculation of $\langle \sigma_z \rangle$ is straightforward; we use an integration region $-L < x < +L$, and we allow $L \rightarrow \infty$. The plane wave regions outside the local spin region then dominate, and, for
real values of $Q$, we obtain
\be
\langle \sigma_z \rangle = {1 \over 18}\bigl\{ 5 - 4 \sqrt{2} {\rm Re} (h_{III}f^\ast_{III} + u_Ir^\ast_I) \bigr\},
\label{sigmaz}
\ee
while, for pure imaginary values of $Q$, the expression for $\langle \sigma_z \rangle$ is somewhat more complicated.

\begin{figure}
\begin{center}
\epsfig{figure=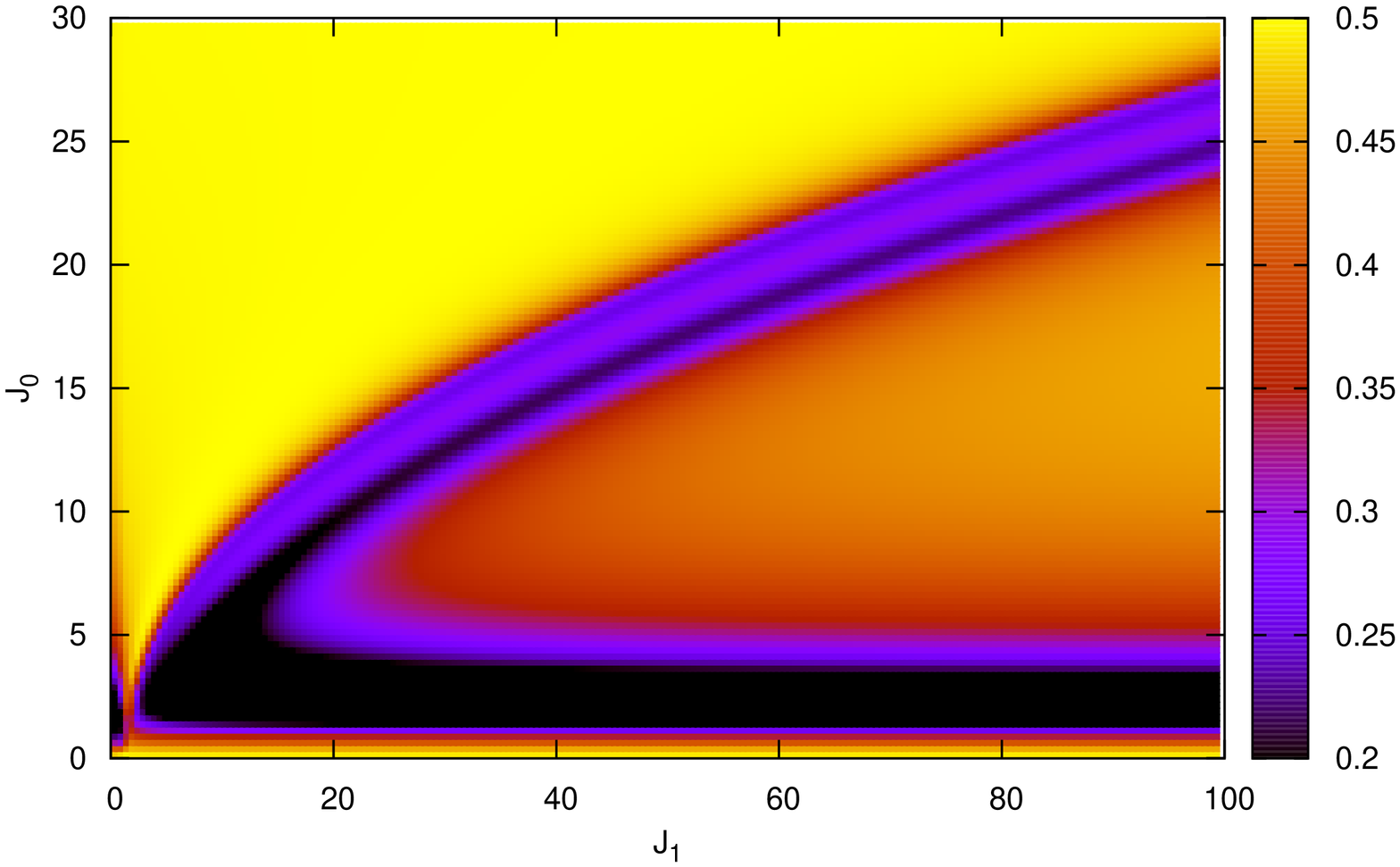,height=2.5in,angle=-90}
\epsfig{figure=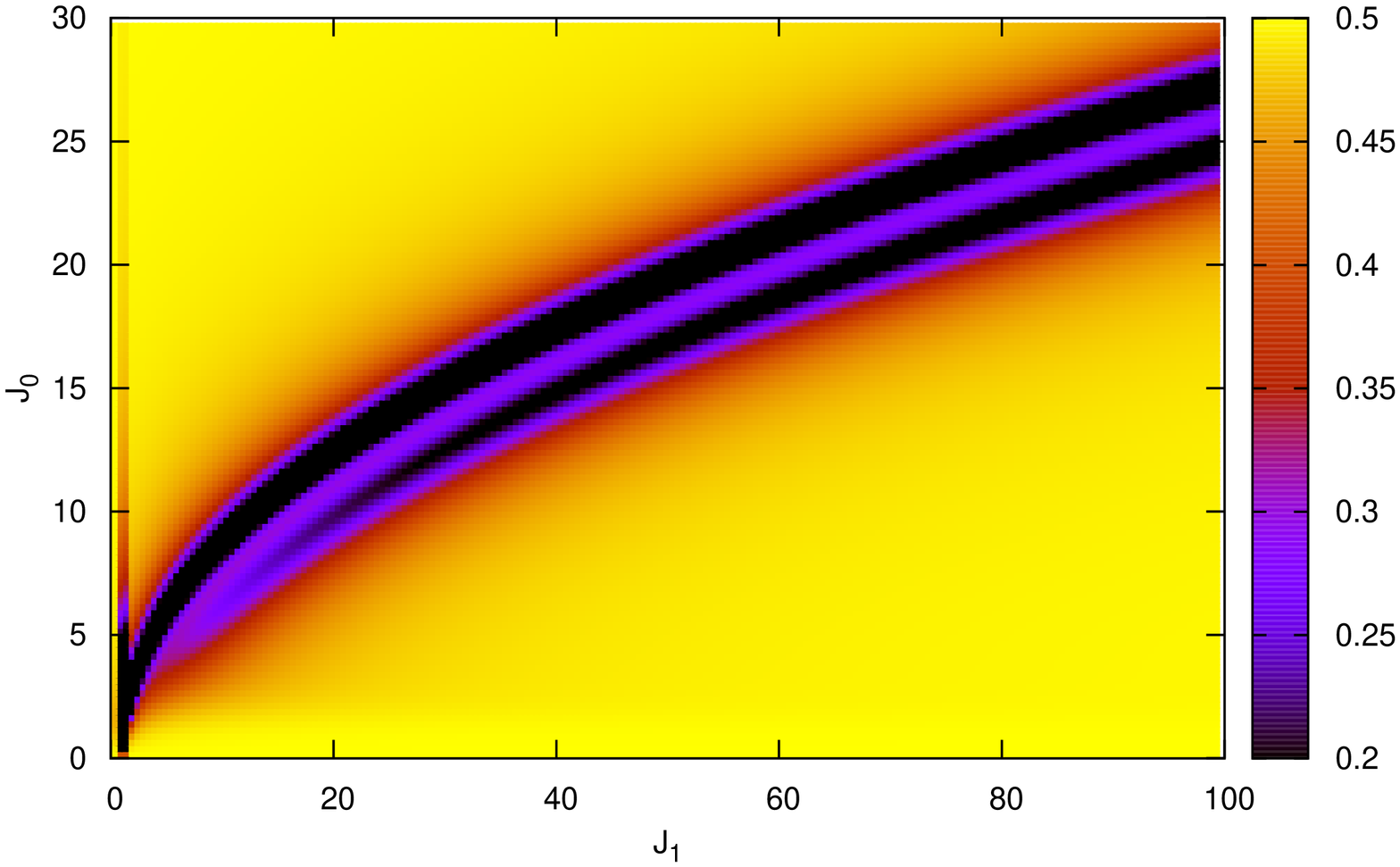,height=2.5in,angle=-90}
\caption{(Color online) (a) Plot of the expectation value of the z-component of the electron spin, $\langle \sigma_z \rangle$, as a function of the two coupling parameters, $J_1$ and $J_0$, based on the plane wave solutions to this problem. The range of $J_1$ and $J_0$ is considerably extended compared to Fig. \ref{fig2sites} to emphasize the presence of the trough (shown in dark color) that extends upwards and to the right. In (b) we show a plot for $1/2 - 0.2*|g_{II}|^2/|f_I|^2$ for the same parameters; the trough is immediately identifiable in this plot, which reinforces our contention that this region of enhanced spin flip scattering is associated with the non-equilibrium bound state (NEBS) represented by $g_{II}$ (a plot of $s_{II}$ yields similar results).}
\label{fig3d_anal}
\end{center}
\end{figure}

\begin{figure}
\begin{center}
\epsfig{figure=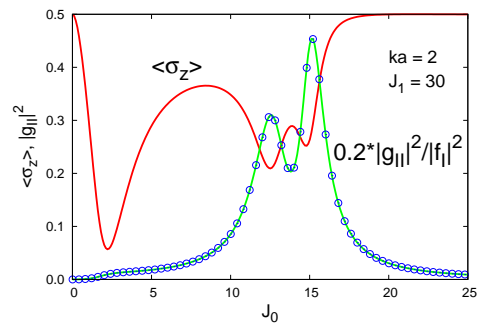,height=2.5in,angle=-90}
\caption{(Color online) (a) Plot of $<\sigma_z>$ (solid (red) curve) vs. $J_0$ for specific values of $J_1$ and $k$, as indicated. Also shown is the coefficient $|g_{II}|^2$ (solid (green) curve), which shows a peak precisely where $<\sigma_z>$ has a significant dip, indicative of enhanced spin flip scattering. Also shown (symbols) is the result of an approximate analytical expression derived in the text. Agreement is extremely good.}
\label{fig_g2}
\end{center}
\end{figure}

In Fig. \ref{fig3d_anal} we show (a) $\langle \sigma_z \rangle$ and (b) $|g_{II}|^2$ as a function of $J_1$ and $J_0$ to
emphasize the connection between the region (described as a `trough') of enhanced spin flip scattering and the non-equilibrium bound state (NEBS). The range of both $J_1$ and $J_0$ is considerably extended compared with Fig. \ref{fig2sites}; nonetheless the qualitative similarities are striking; clearly the analytical solution captures the essence of the numerical one. Furthermore, the analytical approach has allowed us to make the association of the trough of enhanced spin flip scattering with the NEBS. Quantitative details differ, in part because the numerical results are based on a tight-binding model whereas the analytical ones utilize a quadratic dispersion for the itinerant spin. A specific example is given in Fig. \ref{fig_g2}, where both
$\langle \sigma_z \rangle$ and $|g_{II}|^2$ are plotted as a function of $J_0$ (for a specific value of $k$ and $J_1$). Clearly the peaked region in $|g_{II}|^2$ (near $J_0 \approx 15$) corresponds to the dip in $\langle \sigma_z \rangle$, showing strong evidence for the role of the NEBS in enhanced spin flip scattering. For large values of $J_1 >> \epsilon_{\rm el}$, Eq. (\ref{g2}) simplifies somewhat; we get
\be
|g_{II}|^2 = {3 \over 4}\biggl({J_0 \over Q}\biggr)^2 {\sin^2{ka} + (\cos{ka} + 2 v\sin{ka})^2 \over 1 + 4 v^2(\cos{ka} + v\sin{ka})^2},
\label{g2_analytical}
\ee
where $v \approx {J_0 \over k}\bigl[1 - {3 \over 4}{J_0 \over Q}\bigr]$. Similar
analytical expressions can be readily attained for all the coefficients, but they are of limited value.

Eq. (\ref{g2_analytical}) is also plotted in Fig. \ref{fig_g2}, where it is seen to be very accurate (in fact, it is fairly accurate all the way down to $J_1 \approx 2$). The peak region in $|g_{II}|^2$ follows roughly a dispersion relation
\be
J_1 \approx {\epsilon_{\rm el} \over 2} + {9 \over 64} J_0^2,
\label{dispersion}
\ee
and, as has been emphasized already, this corresponds to the region of most intense spin flip scattering (the 'dark trough' region of previous figures). Thus, when $J_0$ and $J_1$ are tuned to satisfy Eq. (\ref{dispersion}) we find an enhanced spin flip process.

\subsection{(d) transmission and reflection amplitudes from the numerical solutions}

\begin{figure}
\begin{center}
\epsfig{figure=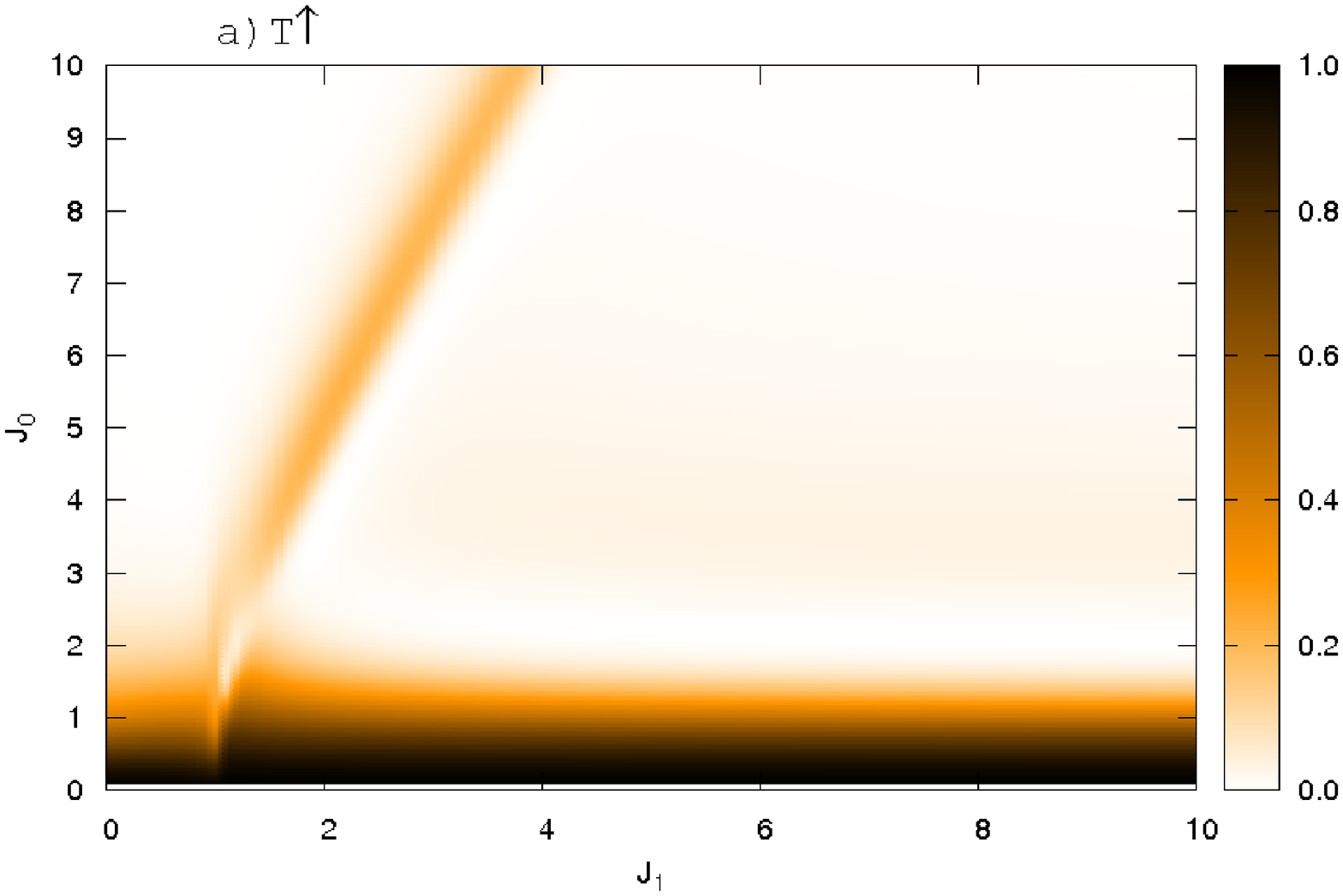,height=3.in,angle=-90}
\epsfig{figure=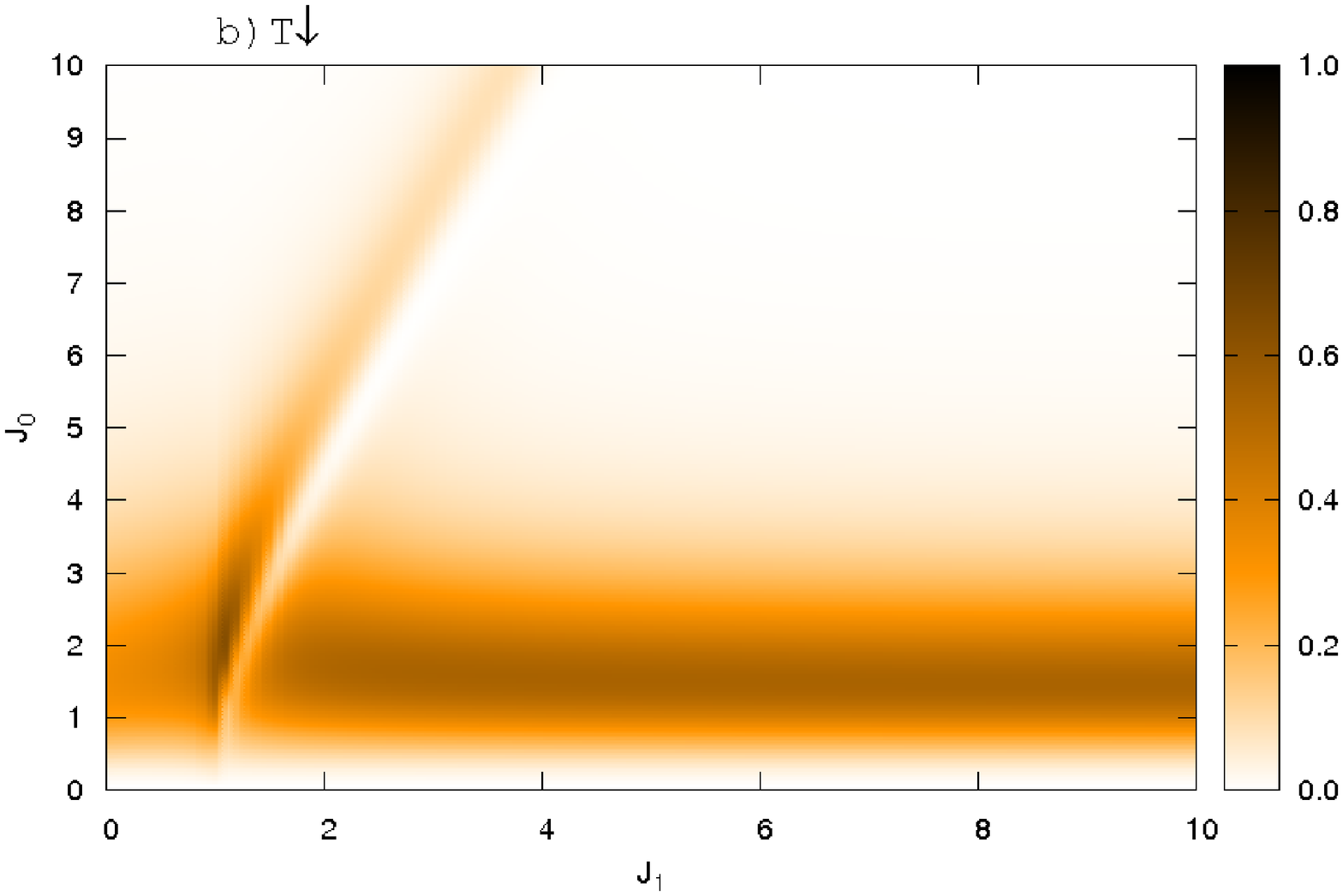,height=3.in,angle=-90}
\epsfig{figure=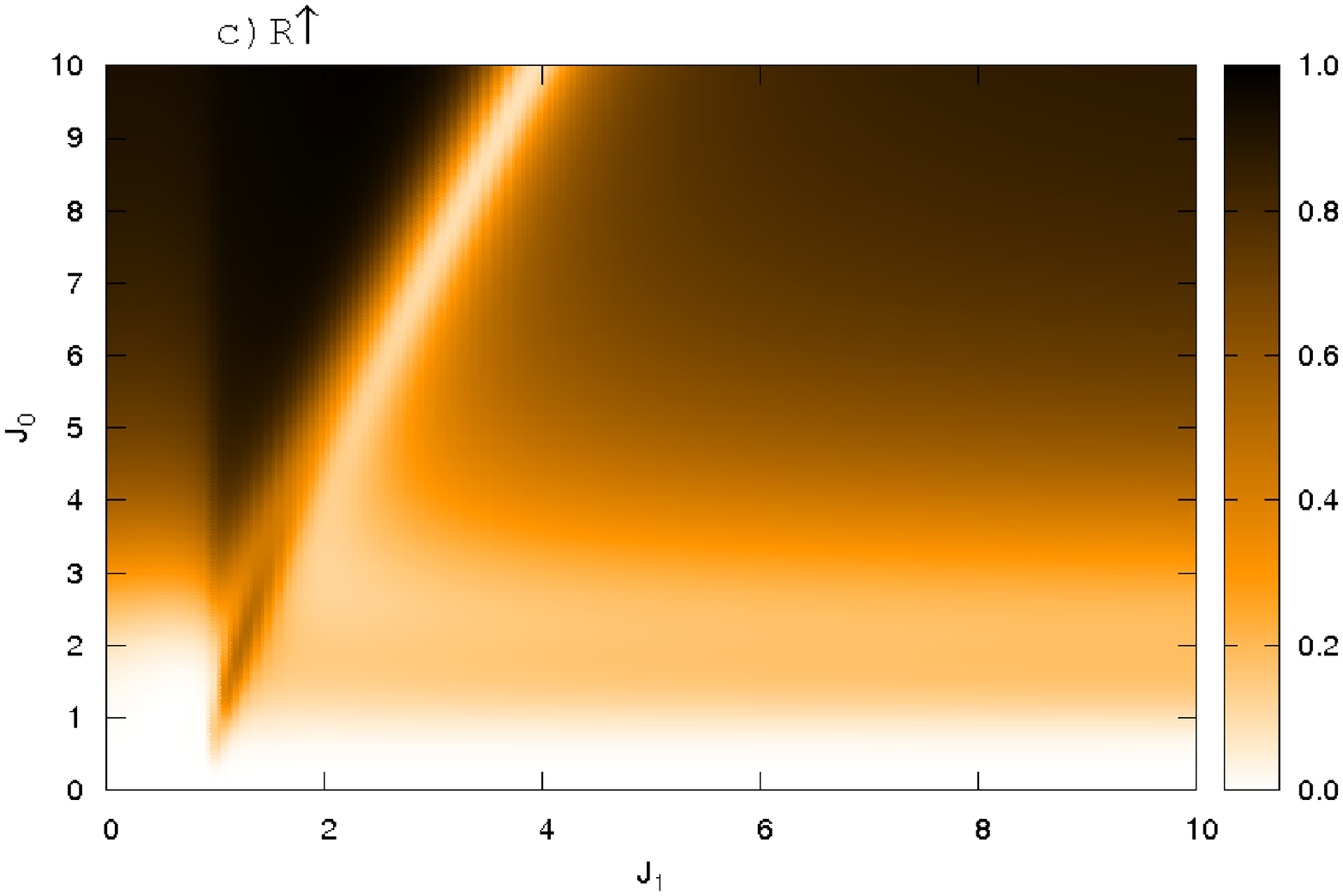,height=3.in,angle=-90}
\epsfig{figure=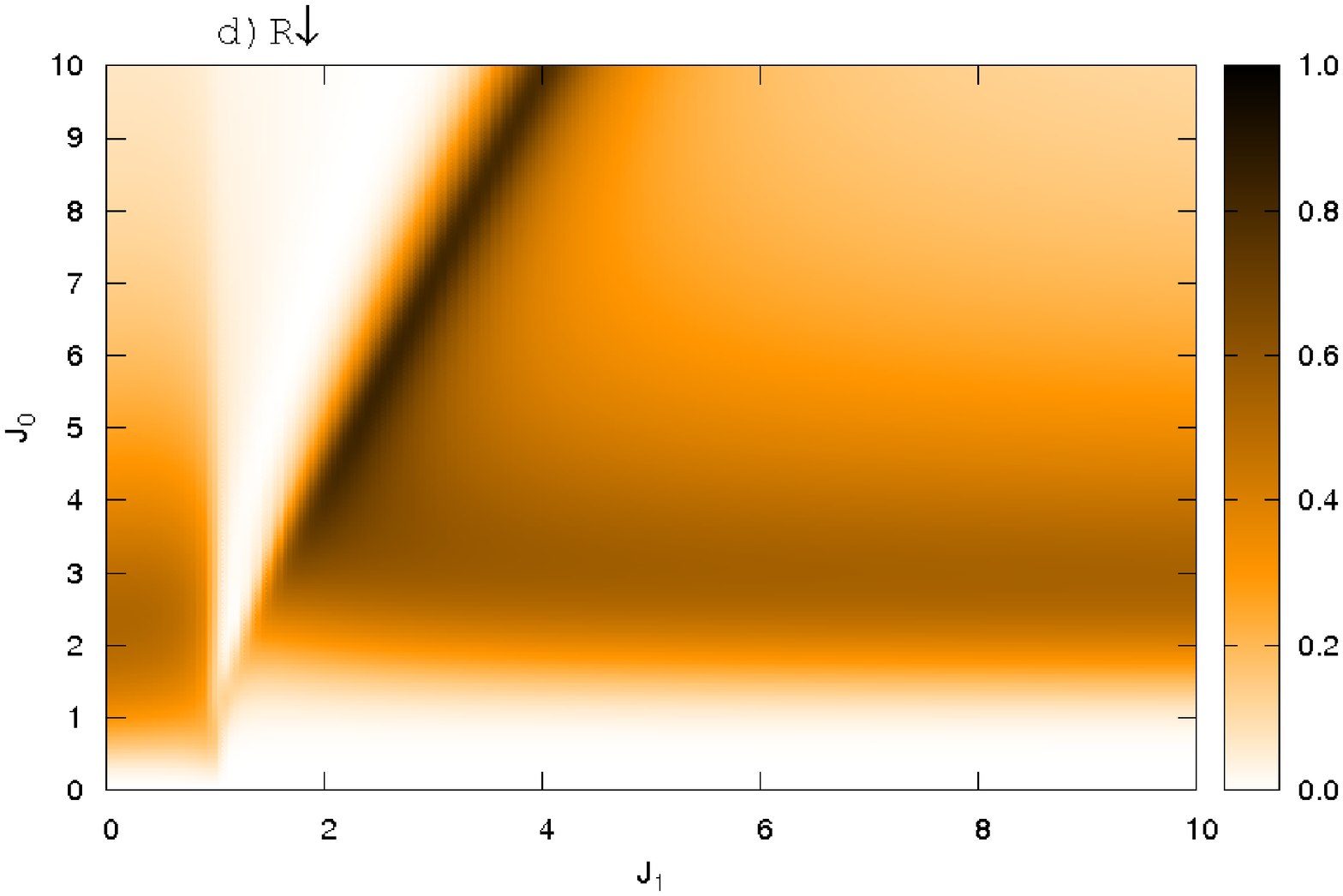,height=3.in,angle=-90}
\caption{(Color online) The (a) transmitted spin-up, (b) transmitted spin-down, (c) reflected spin-up, and (d) reflected spin-down intensities as a function of $J_0$ and $J_1$. These results are obtained with the same conditions as in Fig. \protect\ref{fig2sites}. See the text for further description.}
\label{fig2sites_RT}
\end{center}
\end{figure}

Having established the idea of a NEBS we once again re-examine the numerical solutions.
In particular, one important property from the experimental point of view is that the stationary spins can act as a spin barrier. We have already shown that a large electron-spin interaction ($J_0$)
works as a high potential barrier for the incoming spin. In our framework, for instance,
a large electron-spin interaction acts to prevent any spin-up
component of the electron from transmitting through the stationary spin system. However, in the
$J_0 - J_1$ phase space, at the onset of the trough described, for example, in Fig. \ref{fig2sites}, the transmitted component of the spin-up (and spin-down) electron is enhanced considerably; this is illustrated in the 4 plots shown in Fig. \ref{fig2sites_RT}, where both transmitted and reflected intensities are plotted as a function of $J_0$ and $J_1$. As is clearly evident in (a) and (b), the transmission of both spin species is noticeably enhanced in the trough region. Coincidentally the spin-up reflected component is decreased, while the spin-down reflected component shows an increase. The increase in the transmitted
spin-up component of the electron is not through `direct' transmission. Rather it is achieved through the
spin-flip interaction that generates the component with amplitude $g$ discussed earlier in this section. Recall that in this parameter regime this $g$-component does not exist outside the local spins; it first transforms into the component with amplitude $f$, which represents a propagating wave with both spin-up and spin-down species. These plots therefore reinforce the idea that the electron goes through a two step `virtual' spin flip interaction (creation of the NEBS) in the trough region.

\section{Conclusions}

We have modeled spin current-induced spin torque in the quantum regime with a lattice, on which an itinerant spin (constructed as a wave packet) moves with a kinetic energy given by a tight-binding dispersion, to represent the spin current. Any number of ferromagnetically coupled spins can then be flipped by repeating the process described here with more itinerant electrons, i.e. a current. As described in Ref. \onlinecite{kim07}, this then requires a density matrix description. We have focussed on just two coupled local spins, since this small system contains the essence of the processes we believe are responsible for spin torque: (i) direct spin flip without internal excitation of the local spin system, and (ii) spin flip through inelastic scattering, either real or virtual. The first process exists even for a single local spin, and has been explored previously by us. The second process is the primary subject of this paper, particularly in the regime where, energetically, the itinerant spin becomes momentarily bound in the local system, a phenomenon which we have called the non-equilibrium bound state (NEBS). The description here is for a one dimensional system, but the NEBS should also be present in three dimensions.

An analytical plane wave approach, using a parabolic dispersion for the itinerant spin, helps to elucidate the nature of the spin flip processes. A scattering channel through which a local spin singlet is generated is responsible for the enhanced spin flip scattering along a 'trough' in the $(J_0,J_1)$ phase diagram. This trough is reasonably well described in the plane wave approach by the relation $J_1 = 9J_0^2/64 + \epsilon_{\rm el}/2$.

An experimental observation of the NEBS would be straightforward provided at least one of the parameters $J_0$, $J_1$, or $\epsilon_{\rm el}$ can be tuned in a particular system. In this way the probability of spin flip can be monitored as a function of parameter space, and the NEBS would be identified by a well defined region of enhanced spin flip, corresponding to the 'trough' in Fig. \ref{fig2sites}.

One interesting consequence of our calculation is the possibility of
using the spin chain as an effective spin filter. By tuning the
parameters to correspond to the regimes of enhanced spin flipping, the
spin up electrons will be flipped while the spin down ones will be
unaffected. This effect can be achieved not only for the 2-spin chain but
also for the longer chains, as shown in Fig. \ref{fig20sites}. The resonant `trough'
provides a controllable spin filter through the inter-spin coupling
$J_1$.

\end{document}